\documentclass[twocolumn,english]{revtex4-1}
\usepackage[T1]{fontenc}
\usepackage[latin9]{inputenc}
\usepackage[active]{srcltx}
\usepackage{color}
\usepackage{float}
\usepackage{amsbsy}
\usepackage{amstext}
\usepackage{graphicx}
\usepackage{babel}
\begin{document}
\title{Double frustration and magneto-electro-elastic excitations in collinear
multiferroic materials}
\author{D. C. Cabra}
\affiliation{IFLySiB-CONICET-UNLP and Departamento de F\'{i}sica, Universidad Nacional
de La Plata, 1900 La Plata, Argentina}
\author{A. O. Dobry}
\affiliation{IFIR-CONICET-UNR and Facultad de Ciencias Exactas, Ingeniería y Agrimensura,
Universidad Nacional de Rosario, 2000 Rosario, Argentina}
\author{C. J. Gazza }
\affiliation{IFIR-CONICET-UNR and Facultad de Ciencias Exactas, Ingeniería y Agrimensura,
Universidad Nacional de Rosario, 2000 Rosario, Argentina}
\author{G. L. Rossini }
\affiliation{IFLP-CONICET-UNLP and Departamento de F\'{i}sica, Universidad Nacional
de La Plata, 1900 La Plata, Argentina }
\begin{abstract}
We discuss a model scenario for multiferroic systems of type II (collinear
spins) where the electric dipolar order competes with a frustrated
magnetic order in determining the elastic distortions of the lattice
ion positions. High magnetic frustration due to second neighbors exchange
and small spin easy-axis anisotropy lead to the appearance of the
so called quantum magnetic plateau states. Increasing the magnetic
field above the plateau border produces composite excitations, where
fractionalized spin tertions arise together with spontaneous dipolar
flips (in the form of domain walls) and enhanced localized elastic
distortions. This peculiar magneto-electric effect may be described
by magneto-electric-elastic (MEE) quasiparticles that could be detected
by X-ray and neutron diffraction techniques. Our results are supported
by extensive DMRG computations on the spin sector and self-consistent
equations for the lattice distortions.
\end{abstract}
\maketitle

\section{Introduction\label{sec:Introduction}}

Multiferroic materials, in which magnetic and dipolar order coexist
and interact, are currently under intense investigation. Elastic distortions
of ionic positions, in the form of structural transitions, are deeply
connected with the onset of magnetic and/or ferroelectric order. Some
of these materials do not show signs of spin-orbit correlations, instead
they exhibit collinear spin order and electric polarization below
a common transition temperature; they are called improper type-II
collinear multiferroics {[}\onlinecite{2007-Cheong,2008-vdBrink-Khomskii,2009-Khomskii}{]}.
Several materials with such transitions are described in {[}\onlinecite{key-3,key-4,key-5,key-6,key-7,key-8,key-9,key-10,key-11,key-12,key-13}{]}.

\begin{figure}
\begin{centering}
\includegraphics[scale=0.6]{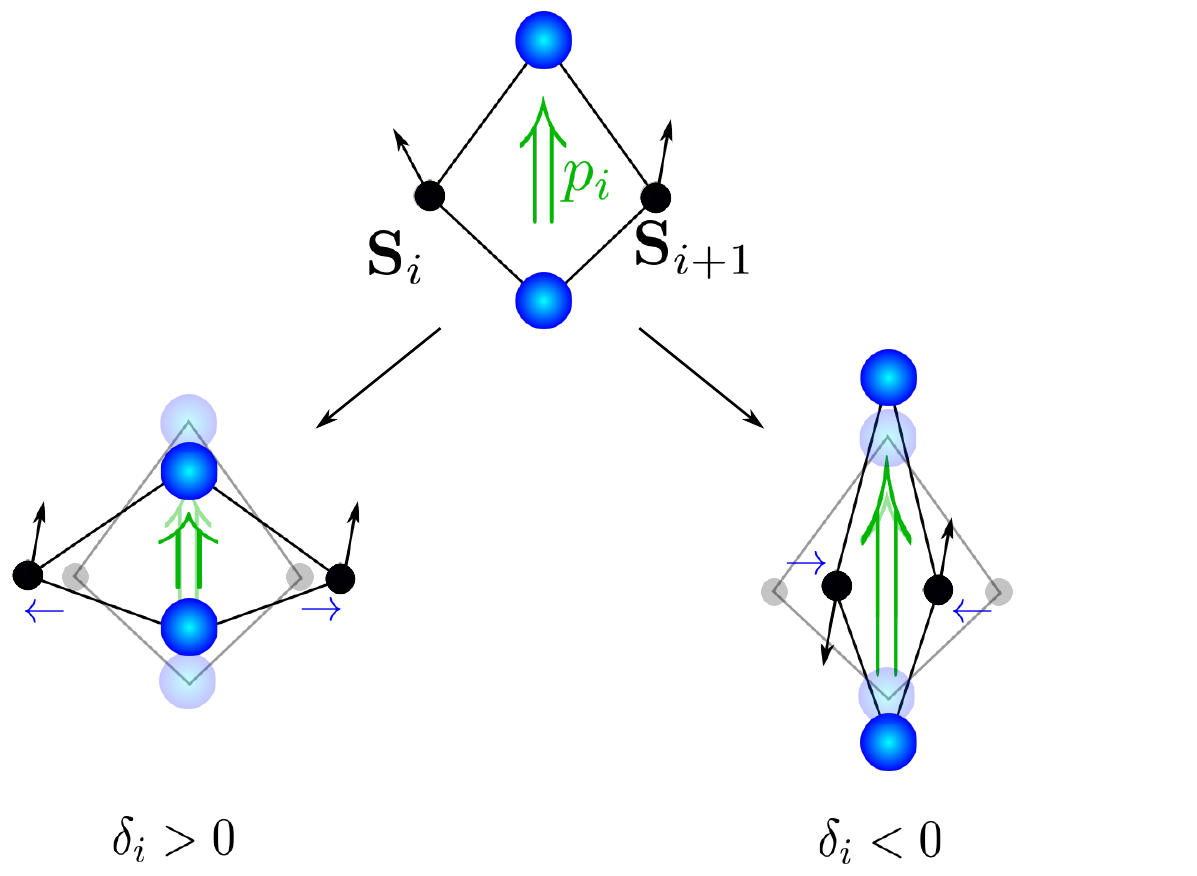}
\par\end{centering}
\caption{Schematic picture for the pantograph mechanism coupling electric dipoles
to the lattice. Black dots represent magnetic sites and blue spheres
represent a charge distribution giving rise to dipolar moments. Green
double arrows represent these dipolar moments that might point up
or down. Displacements of magnetic sites, indicated by blue arrows,
produce lattice bond length distortions $\delta_{i}$ that modify
the strength of local dipoles.\textcolor{red}{{} \label{fig:Schematic-pantograph}}}
\end{figure}

We have recently proposed a microscopic model in {[}\onlinecite{2021-pantograph-II}{]}
(cited as Ref. I in the following) describing the interaction between
spins and electric dipoles in a quasi one-dimensional multiferroic
system via elastic distortions. A key ingredient of this model, shown
in Figure \ref{fig:Schematic-pantograph}, is the variation of dipolar
moments according to the distance between neighboring spin sites;
as dipoles are shortened when spin separation is increased, the model
reminds a pantograph mechanism. Thus the dipole-dipole interaction
produces a back reaction of the dipolar degrees of freedom on the
elastic distortions, and consequently in the magnetic sector. A salient
feature is the observation of an ordered dipolar phase with period
three, which shows up in the presence of an appropriate homogeneous
external electric field (see Figure 3 in Ref. I) because of the long
range character of dipolar interactions (c.f. {[}\onlinecite{2019-pantograph-I}{]}
where only nearest neighbors dipolar interactions are discussed).
Some properties of this dipolar phase, denoted as $\Uparrow\Uparrow\Downarrow$
in the following {[}\onlinecite{footnote-1}{]}, motivate the present
work.

We investigate here the commensurability interplay of the $P=1/3$
period three dipolar order $\Uparrow\Uparrow\Downarrow$ (see Figure
\ref{fig: simple patterns}-C below) with the period three magnetic
configurations observed in many frustrated magnetic materials with
$M=1/3$ magnetization plateaus ($M$ expressed as a fraction of saturation).
In most studies the $M=1/3$ plateau state is found to form a collinear
$\uparrow\uparrow\downarrow$ (see Fig \ref{fig: simple patterns}-B)
classical pattern {[}\onlinecite{2003-Okunishi-a}{]}, but a quantum
$\bullet\!\!-\!\!\bullet\uparrow$ order (where $\bullet\!\!-\!\!\bullet$
stands for a spin singlet, see Figure \ref{fig: simple patterns}-A)
has been also predicted for spin $S=1/2$ modulated isotropic Heisenberg
chains {[}\onlinecite{2005-Hida-Affleck}{]}. The robustness of magnetic
plateau states, given by an energy gap in the magnetization spectrum,
makes them good candidates for technological applications. Relatedly,
the characteristics of the elementary $\Delta S^{z}=1$ excitations
at the high field border of the plateaus deserve detailed analysis.

We find that the dipolar order introduced by the long range dipolar
interactions indeed competes with the magnetic order set by the magnetic
frustration at the M=1/3 plateau, opening a non trivial scenario which
we dub \emph{double frustration}. Our analysis predicts unusual effects
due to this scenario. For low anisotropy and high magnetic frustration,
favoring quantum fluctuations, the double frustration stabilizes a
quantum structure at the $M=1/3$ plateau state. In contrast, either
for higher easy-axis anisotropy or for lower magnetic frustration,
or both, the double frustration competition leads to the spontaneous
parity symmetry breaking of the classical collinear $M=1/3$ plateau
state. We must stress that, in a more general situation with charge
order along the chain, parity breaking would imply the appearance
of longitudinal electric polarization {[}\onlinecite{2008-vdBrink-Khomskii}{]}.
These results are clear signals of the role of dipolar interactions
and may guide the search for materials realizing strong magneto-electric
effects. 

After analyzing the plateau states, we discuss the excitations caused
by the increase of the magnetic field. We show that the $\Delta S^{z}=1$
magnon on top of the $M=1/3$ state fractionalizes into three $S^{z}=1/3$
spatially separated solitons, where elastic distortions adapt to the
magnetic order. This change in the distortion pattern induces, in
the dipolar sector, a spontaneous unit polarization change which in
turn fractionalizes into three sharp domain walls. This emergent magneto-electric
effect, that is the polarization change induced by a magnetic field
mediated by elastic distortions, is one of the main results in the
present paper. 

Both the nature of the plateau state structure and the appearance
of intertwined magnetic and electric fractional excitations, mediated
by the lattice, are experimentally accessible by neutron scattering
for the spin-channel and by X-ray scattering for the lattice distortions. 

The paper is organized as follows: in Section II we describe the microscopic
degrees of freedom, the Hamiltonian and the parameters that define
our model scenario. We also discuss the computational method. In Section
III we describe the double frustration effect, that is a competence
of the elastic order convenient to magnetic frustration with the elastic
order driven by dipolar interactions, stressing the differences between
classical and quantum regimes. We devote Section IV to describe the
effects of increasing the magnetic field, which leads to fractional
magnetic and dipolar excitations as well as localized elastic domain
walls with clear different behavior with respect to the classical
or quantum plateau state structure. Conclusions are presented in Section
V.

\medskip{}

\section{The model scenario\label{sec:The-model}}

We briefly summarize here the model proposed in Ref. I. It describes
spin $S=1/2$ magnetic atoms $\boldsymbol{S}_{i}$ in a linear chain
with elastic degrees of freedom $\delta_{i}$ describing the variation
of the bond length between neighboring magnetic atoms $\boldsymbol{S}_{i}$
and $\boldsymbol{S}_{i+1}$, from a regular lattice constant length
$a$ to distorted lengths $a+\delta_{i}$. Electric dipolar moments
$p_{i}$, normal to the chain direction, are located amid magnetic
atoms $\boldsymbol{S}_{i}$ and $\boldsymbol{S}_{i+1}$; when the
magnetic lattice is distorted the distance between dipoles $p_{i}$
and $p_{i+1}$ also changes from $a$ to $a+\eta_{i}$, with $\eta_{i}=\left(\delta_{i}+\delta_{i+1}\right)/2$.
A fixed chain length condition is assumed, imposing a null constraint
on the sum of local distortions. 

In order to describe the several materials mentioned in the Introduction,
the model includes easy-axis anisotropic antiferromagnetic interactions
$J_{1}$ and $J_{2}$ between nearest (NN) and next-to-nearest (NNN)
neighbors respectively, which produce the magnetic frustration. Electric
dipoles interact via long range Coulomb interactions. Both the magnetic
and dipolar sectors are coupled to the lattice in the most natural
way. The spin sector does it by a standard, spin-Peierls type, distance
dependence of the NN exchange coupling linearly expanded as $J_{1}(i,i+1)=J_{1}\left(1-\alpha\delta_{i}\right)$.
On the other hand, the dipolar sector couples to elastic distortions
through the $1/r^{3}$ distance dependence of long range dipole-dipole
interactions and by the geometric mechanism mentioned in the Introduction,
affecting the charge distribution in-between magnetic atoms as they
are displaced. This so called pantograph effect is modeled by a linear
expansion of the dipole moments $p_{i}=p_{0}\left(1-\beta\delta_{i}\right)\sigma_{i}$
where $p_{0}$ is the undistorted dipole magnitude, $\beta$ measures
the pantograph electro-elastic coupling and $\sigma_{i}=\pm1$ is
the Ising variable describing dipole orientations. The elastic distortions
have an energy cost given by a stiffness constant $K$, providing
an indirect magneto-electric coupling. More details on the model can
be found in Ref. I.

The complete Hamiltonian reads

\begin{eqnarray}
H & = & \sum_{i}\left[J_{1}\left(1-\alpha\delta_{i}\right)\left(\mathbf{S}_{i}\cdot\mathbf{S}_{i+1}\right)_{\Delta}+J_{2}\left(\mathbf{S}_{i}\cdot\mathbf{S}_{i+2}\right)_{\Delta}\right]\nonumber \\
 & + & \frac{K}{2}\sum_{i}\delta_{i}^{2}+\lambda_{dip}\sum_{i<j}\frac{1}{r_{ij}^{3}}p_{i}p_{j}\label{eq: H-short}
\end{eqnarray}
where $\left(\mathbf{S}_{i}\cdot\mathbf{S}_{j}\right)_{\Delta}$ stands
for the anisotropic spin-spin product, $\lambda_{dip}>0$ is the Coulomb
coupling constant and $r_{ij}$ is the distance between dipoles $p_{i}$
and $p_{j}$, which depends on distortions. With the previous notation,
expanding the dipolar $1/r^{3}$ dependence up to linear order in
distortions and truncating up to second neighbours (see Ref. I), the
explicit model Hamiltonian is given by \begin{widetext}
\begin{eqnarray}
H & = & \sum_{i}\frac{J_{1}}{\Delta}\left(1-\alpha\delta_{i}\right)\left(S_{i}^{x}S_{i+1}^{x}+S_{i}^{y}S_{i+1}^{y}+\Delta S_{i}^{z}S_{i+1}^{z}\right)+\sum_{i}\frac{J_{2}}{\Delta}\left(S_{i}^{x}S_{i+2}^{x}+S_{i}^{y}S_{i+2}^{y}+\Delta S_{i}^{z}S_{i+2}^{z}\right)\nonumber \\
 & + & \frac{K}{2}\sum_{i}\delta_{i}^{2}\\
 & + & J_{e}\sum_{i}\left(\sigma_{i}\sigma_{i+1}+\frac{1}{8}\sigma_{i}\sigma_{i+2}\right)\label{eq: H-full-expanded}\\
 & - & J_{e}\sum_{i}\delta_{i}\left[\left(\beta+\frac{3}{2a}\right)\left(\sigma_{i-1}\sigma_{i}+\sigma_{i}\sigma_{i+1}\right)+\frac{1}{8}\left(\beta+\frac{3}{4a}\right)\left(\sigma_{i-2}\sigma_{i}+\sigma_{i}\sigma_{i+2}\right)+\frac{3}{16a}\sigma_{i-1}\sigma_{i+1}\right]\nonumber 
\end{eqnarray}
\end{widetext}where $\Delta\geq1$ is the easy-axis anisotropy and
$J_{e}=\lambda_{dip}p_{0}^{2}/a^{3}$ is the effective dipolar interaction
coupling, ($J_{1}$, $J_{2}$ and $J_{e}$ in energy units). Magnetic
couplings $J_{1}$ and $J_{2}$ are divided by $\Delta$ in order
to reach the Ising regime in the highly anisotropic $\Delta\to\infty$
limit. 

Following Ref. I we are interested on a parameter region where the
magnetic and dipolar couplings are of the same order of magnitude,
so both the spin and dipole configurations are relevant to determine
the ground state of the system. Also the magneto-elastic coupling
$\alpha$ and the electro-elastic coupling $\beta$ are similar, in
order to provide an efficient elastically mediated magneto-electric
interaction. We then avoid the multiplicity of parameters in the Hamiltonian
(\ref{eq: H-full-expanded}) by taking $Ka^{2}$ as the energy unit
and fixing $J_{1}$, $J_{e}$, $\alpha$ and $\beta$ at convenient
values detailed below. Only $J_{2}$ and $\Delta$ will be varied
to explore the incidence of magnetic frustration and easy-axis anisotropy
(measured as $\gamma=1/\Delta$ in Ref. I) in the ground state properties
of the system. Different values of $J_{1}$, $J_{e}$, and $\beta$
can be studied similarly.

External electric and magnetic fields $\boldsymbol{E}$ and $\boldsymbol{B}$
will be adjusted to drive the system to the peculiar double frustration
scenario we discuss in the present paper. This is the region where
the electric field polarizes the otherwise antiferroelectric dipolar
sector (driven by $J_{e}$) up to $1/3$ of saturation, provoking
the period three $\Uparrow\Uparrow\Downarrow$ dipolar pattern (see
Figure 3 in Ref. I) and the magnetic field sets the spin degrees of
freedom in the $M=1/3$ plateau region (see Figure 6 in Ref. I). For
a magneto-elastic, spin-Peierls, chain (not coupled to electric dipoles),
this plateau is known to appear together with an energetically favorable
period three elastic distortion {[}\onlinecite{2006-Vekua-etal,2007-Gazza-etal,2007-Rosales-et-me}{]}.
On the other hand, for the electro-elastic chain obtained from the
Hamiltonian (\ref{eq: H-full-expanded}) when the spin sector is decoupled
($\alpha=0$), the $\Uparrow\Uparrow\Downarrow$ dipolar pattern also
comes along with period three elastic distortions (as discussed in
Ref. I) bringing closer (farther) antiparallel (parallel) dipoles.
This situation might be reversed, for instance in the presence of
itinerant electrons, since they may induce RRKY-like interactions
between dipoles leading to ferroelectric effective couplings (see
e.g. {[}\onlinecite{2019-Jin}{]}).

The question arises whether the elastic distortions compete or collaborate
in lowering the ground state energy of the magneto-electro-elastic
multiferroic system. We show below that they do compete, with profound
consequences both in the magnetic plateau configuration and the magnetic
excitations at the high field border of the plateau.

Our results are based on extensive numerical computations following
an iterative self-consistent method {[}\onlinecite{1997-Feiguin}{]}
where the spin sector is solved exactly by Density Matrix Renormalization
Group (DMRG) techniques {[}\onlinecite{1992-White}{]}. At each iteration,
for a given configuration of dipoles $\{\sigma_{i}\}$ and a quantum
state for the spins $\{\boldsymbol{S}_{i}\}$, the lattice distortions
$\delta_{i}$ are obtained by minimizing the elastic energy under
a fixed chain length condition. Unconstrained distortions $\delta_{i}^{\text{free}}$
are computed through the local zero gradient conditions
\begin{eqnarray}
K\delta_{i}^{\text{free}} & = & \alpha J_{1}/\Delta\langle\left(S_{i}^{x}S_{i+1}^{x}+S_{i}^{y}S_{i+1}^{y}+\Delta S_{i}^{z}S_{i+1}^{z}\right)\rangle\nonumber \\
 & + & J_{e}\left(\beta+\frac{3}{2a}\right)\left(\sigma_{i-1}\sigma_{i}+\sigma_{i}\sigma_{i+1}\right)\nonumber \\
 & + & \frac{1}{8}J_{e}\left(\beta+\frac{3}{4a}\right)\left(\sigma_{i-2}\sigma_{i}+\sigma_{i}\sigma_{i+2}\right)\nonumber \\
 & + & J_{e}\frac{3}{16a}\sigma_{i-1}\sigma_{i+1}-\beta\varepsilon\sigma_{i},\label{eq: self consistence}
\end{eqnarray}
where $\epsilon=2p_{0}E$ is the normalized electric field, while
the constraint is imposed as 
\begin{equation}
\delta_{i}=\delta_{i}^{\text{free}}-\overline{\delta^{\text{free}}}.\label{eq: constraint}
\end{equation}
where the bar stands for average value along the chain. Interestingly,
the self-consistent conditions in Eqs. (\ref{eq: self consistence},
\ref{eq: constraint}) also allow for a qualitative analysis of the
influence of spin-spin and dipole-dipole correlations on the elastic
distortions.

\section{Double frustration effect\label{sec:Double-frustration}}

\subsection{Qualitative description}

The elastic distortions associated with the $M=1/3$ magnetic plateau
configuration, and those associated with the $\Uparrow\Uparrow\Downarrow$
dipolar pattern, can be qualitatively described considering the nearest
neighbor (NN) interactions in Eq. (\ref{eq: self consistence},\ref{eq: constraint}).
We then provide the numerical evidence for the outcoming picture in
the following subsection.

$M=1/3$ magnetic plateaus come in two flavors, dubbed classical and
quantum {[}\onlinecite{2005-Hida-Affleck}{]}. In the so called classical
plateau spin components parallel to the magnetic field have non vanishing
$\langle S_{i}^{z}\rangle$ expectation value in an ordered pattern
with two positive, one negative terms that we represent by $\uparrow\uparrow\downarrow$.
These expectation values are reduced by quantum fluctuations in the
isotropic $\Delta=1$ case, but approach $\pm0.5$ in the highly easy-axis
anisotropic case $\Delta\gg1$. Spin-spin correlations $\langle\boldsymbol{S}_{i}\cdot\boldsymbol{S}_{i+1}\rangle$
are positive between ferromagnetic (parallel) neighbors $\uparrow\uparrow$
and negative between antiferromagnetic (antiparallel) neighbors $\uparrow\downarrow$
and $\downarrow\uparrow$, approaching the Ising correlations $\pm0.25$
for $\Delta\gg1$. From Eq. (\ref{eq: self consistence}), the correlation
$\langle\boldsymbol{S}_{i}\cdot\boldsymbol{S}_{i+1}\rangle$ affects
the bond distortion $\delta_{i}$; the $\uparrow\uparrow\downarrow$
spin configuration favors distorted long bonds between ferromagnetic
neighbors and short bonds between antiferromagnetic neighbors, that
is a \textquotedbl long-short-short\textquotedbl{} (L-S-S) distortion
pattern (see Figures \ref{fig:Schematic-pantograph} and \ref{fig: simple patterns}-A).
Notice that the antiferromagnetic coupling $J_{1}\left(1-\alpha\delta_{i}\right)$
gets stronger for \textquotedbl satisfied\textquotedbl{} antiferromagnetically
aligned neighbors and weaker for \textquotedbl frustrated\textquotedbl{}
ferromagnetically aligned neighbors. 

In contrast, in the so called quantum plateau two neighboring spins
(out of three) tend to form singlets while the third one points up,
in a configuration that we represent by $\bullet\!\!-\!\!\bullet\uparrow$
(see Figure \ref{fig: simple patterns}-B). In an ideal case the spins
forming a quantum singlet would have $\langle S_{i}^{z}\rangle=0$
and the third one $\langle S_{i}^{z}\rangle=0.5$, with singlet correlation
$\langle\boldsymbol{S}_{i}\cdot\boldsymbol{S}_{i+1}\rangle=-0.75$
and vanishing correlation between the spin up and its neighbors; the
real situation may be characterized as a quantum plateau when the
spin expectation and spin-spin correlation values show a tendency
to such pattern. Again from Eq. (\ref{eq: self consistence}) one
can see that a very negative singlet-like correlation strongly favors
a short bond at the expense of long bonds (according to Eq. (\ref{eq: constraint}))
where spin correlations are close to zero, giving rise to a \textquotedbl short-long-long\textquotedbl{}
(S-L-L) distortion pattern. Notice that the singlets are more likely
to appear in the isotropic case $\Delta=1$, while the easy-axis anisotropy
$\Delta>1$ diminishes transverse correlations and favors the classical
configuration. 

In turn, the NN dipolar correlations are related to lattice distortions
through the second line of Eq. (\ref{eq: self consistence}): bond
distortion $\delta_{i}$ is influenced by the correlations of the
dipole $\sigma_{i}$ located at the bond $i$ with NN dipoles at both
sides. The $\Uparrow\Uparrow\Downarrow$ configuration then favors
short bonds where the dipole $\Downarrow$ is located, at the expense
of generating long bonds where the dipoles point $\Uparrow$ to fulfill
the constraint in Eq. (\ref{eq: constraint}), preferring to induce
a S-L-L distortion pattern (see Figure \ref{fig: simple patterns}-C).
Recalling that dipoles remain always midway between adjacent magnetic
atoms, in terms of dipole positions these magnetic lattice distortions
make antiparallel dipoles get closer, and parallel dipoles get further
away. 
\begin{center}
\begin{figure}
\begin{centering}
\includegraphics[scale=0.5]{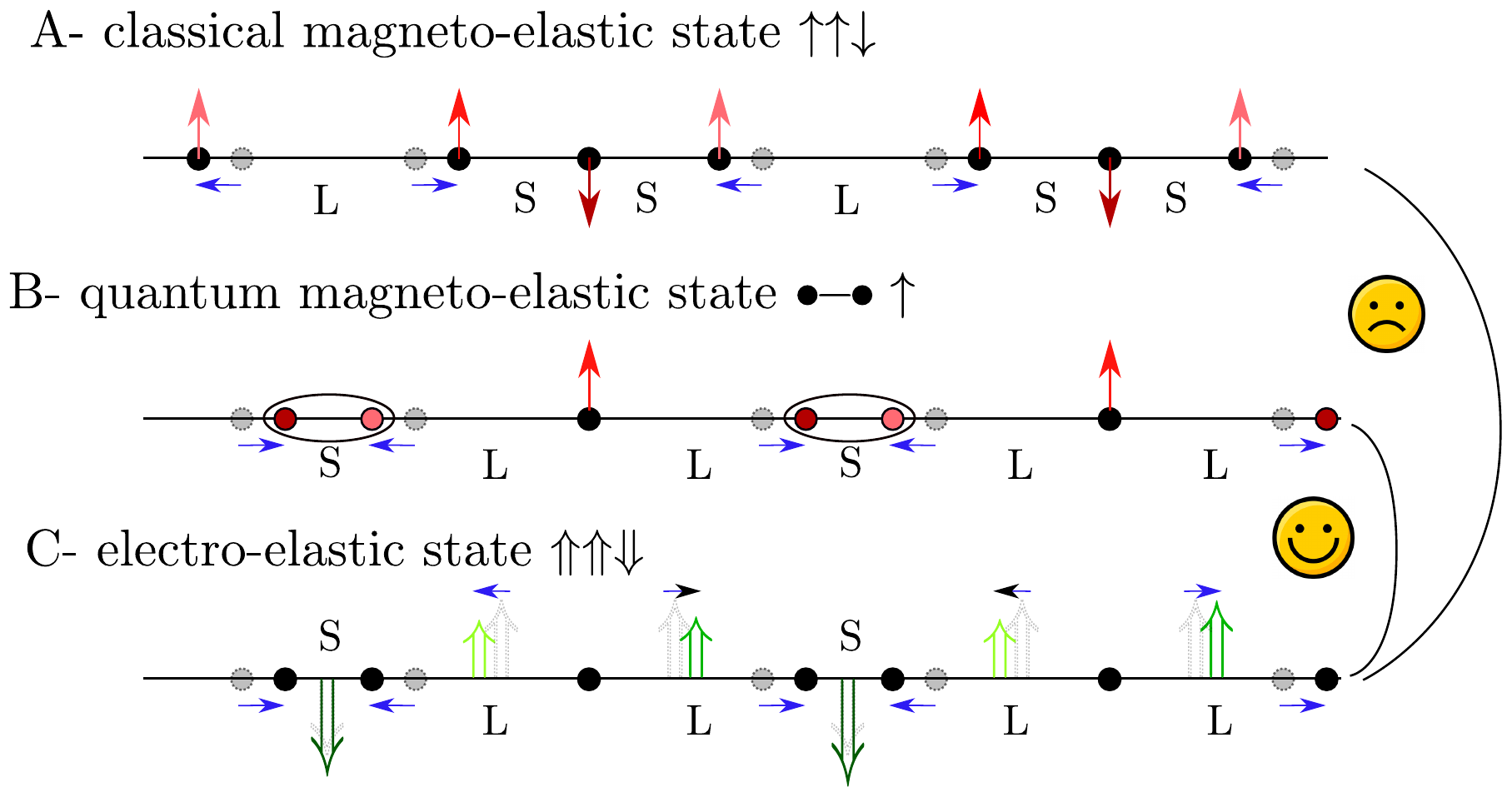}
\par\end{centering}

\caption{\label{fig: simple patterns}Qualitative picture of magneto-elastic
distortions at the classical $\uparrow\uparrow\downarrow$ and the
quantum $\bullet\!\!-\!\!\bullet\uparrow$ magnetic configurations
discussed in {[}\onlinecite{2005-Hida-Affleck}{]} (A and B respectively),
and a picture of the electro-elastic distortions at the dipolar $\Uparrow\Uparrow\Downarrow$
configuration (C) found in Ref. I. Spins are represented by single
arrows in tones of red, dipoles by double arrows in tones of green.
Three color tones (light, medium, dark) are used to facilitate the
recognition of variables every three sites, in relation to data given
below. Ionic and dipole displacements are indicated by blue arrows
(non distorted positions indicated with gray faded symbols). The size
change of dipoles in C is demonstrative of the pantograph mechanism
effect. In panel A the bond length follows a \textquotedbl long-short-short\textquotedbl{}
pattern (L-S-S in the figure), providing a magnetic energy gain by
getting closer (farther) antiparallel (parallel) spins. In panel B
the magnetic energy gain is obtained by tightening singlet bonds following
a \textquotedbl short-long-long\textquotedbl{} pattern (S-L-L in
the figure). In panel C a dipolar energy gain stems from getting closer
(farther) antiparallel (parallel) dipoles, with the same S-L-L bond
distortions as in panel B. Electro-elastic distortions in C are compatible
with the quantum magneto-elastic distortions in B, but not with the
classical ones in A.}
\end{figure}
\par\end{center}

From this qualitative discussion, the electro-elastic dipolar configuration
$\Uparrow\Uparrow\Downarrow$ found in Ref. I is compatible with the
quantum magnetic plateau configuration but competes with the classical
plateau configuration, which is usually the one observed in homogeneous
$J_{1}-J_{2}$ magnetically frustrated spin chains in a wide variety
of regimes (isotropic with {[}\onlinecite{2006-Vekua-etal,2007-Gazza-etal}{]}
and without {[}\onlinecite{2003-Okunishi-a}{]} elastic coupling,
anisotropic {[}\onlinecite{2003-Okunishi-b}{]}). Then, the coupling
to dipolar degrees of freedom through lattice distortions introduces
a second frustration mechanism. Our numerical analysis below provides
clear surprising effects due to this double frustration scenario.
In a regime of low anisotropy and high magnetic frustration, favoring
quantum fluctuations, this second frustration is responsible for the
stabilization of a quantum $M=1/3$ plateau state. In contrast, for
higher easy-axis anisotropy and/or lower magnetic frustration, the
second frustration competition leads to a spontaneous parity symmetry
breaking in the classical $M=1/3$ plateau state.\textcolor{red}{{} }

\subsection{Numerical DMRG analysis}

We have performed an extensive numerical computation of the ground
state of the model in Eq. (\ref{eq: H-full-expanded}), in the presence
of magnetic and electric fields driving the system to magnetization
$M=1/3$ and polarization $P=1/3$. In order to evaluate the role
of magnetic frustration and easy-axis anisotropy we explored the $J_{2}/J_{1}-\Delta$
plane, fixing the remaining parameters at $J_{1}=0.5Ka^{2}$, $J_{e}=0.2Ka^{2}$
and $\alpha=\beta=0.2$ with $K=a=1$; correspondingly the electric
field is taken as $\epsilon=0.16$ (see Figure 3 in Ref. I).

The ground state is found through an iterative numerical analysis
based on DMRG to solve the magnetic sector in the adiabatic Eq. (\ref{eq: self consistence}),
along the lines stated in {[}\onlinecite{1997-Feiguin}{]} and implemented
in a similar context in  {[}\onlinecite{2019-pantograph-I,2021-pantograph-II}{]}.
At each point chosen in the $J_{2}/J_{1}-\Delta$ plane the ground
state of the system is found as follows: starting from the $\delta_{i}$
and $\sigma_{i}$ configuration that solves the electro-elastic part
of the Hamiltonian, the quantum ground state of the spin system is
obtained by the DMRG algorithm. Therefore, we re-obtain the set of
$\delta_{i}$ from Eqs. (\ref{eq: self consistence},\ref{eq: constraint})
and prove different $\sigma_{i}$ to minimize the total energy, until
convergence. We use periodic boundary conditions, and we have kept
the truncation error less than $O(10^{-12})$, during up to more than
100 sweeps in the worst cases. This assures that errors of the DMRG
computation are smaller than symbol sizes in each figure. The DMRG
computations were implemented using the ITensor software library {[}\onlinecite{ITensor}{]}.

\begin{figure}
\centering{}\includegraphics[scale=0.7]{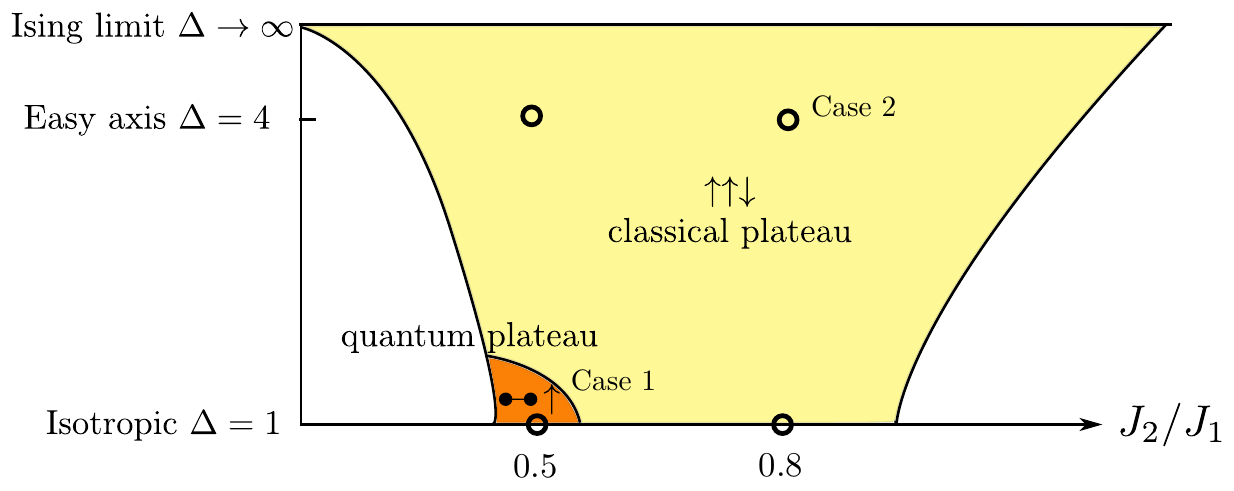}\caption{\label{fig: phase-diagram}Schematic phase diagram in the frustration
ratio ($J_{2}/J_{1}$) and the easy-axis anisotropy ($\Delta$) plane.
The colored regions indicate the parameter regimes where $M=1/3$
plateaus are observed in  magnetization curves. The robust magnetic
order giving rise to the plateau is mostly a collinear $\uparrow\uparrow\downarrow$
classical structure (yellow region) but turns into a quantum $\bullet\!\!-\!\!\bullet\uparrow$
state (orange region) when high frustration and low anisotropy enhance
the quantum fluctuations. The circles mark the four points detailed
in this work (unnumbered ones in Supplemental Material {[}\onlinecite{SM}{]}).}
\end{figure}

We have covered a wide region of the $J_{2}/J_{1}-\Delta$ plane.
From this exploration, we found distinct regimes that we describe
below. We paid attention to the isotropic case $\Delta=1$, mainly
for theoretical reasons, and to high values of $\Delta$ where one
expects a classical behavior which may be in closer relation to real
materials. Regarding the frustration ratio $J_{2}/J_{1}$, we distinguish
moderate and highly frustrated values (see Figure 6 in Ref. I). Representative
selected points are:
\begin{itemize}
\item $\Delta=1$, $J_{2}/J_{1}=0.5$. Due to the isotropic Heisenberg interaction
and the high magnetic frustration ($J_{2}/J_{1}=0.5$ is the maximally
frustrated point in the case of Ising interactions) quantum fluctuations
are enhanced at this point. 
\item $\Delta=4$, $J_{2}/J_{1}=0.8$. Easy-axis anisotropy and low magnetic
frustration inhibit quantum fluctuations, favoring classical behavior.
\item $\Delta=1$, $J_{2}/J_{1}=0.8$. Selected as a point with isotropic
Heisenberg interaction and low frustration. 
\item $\Delta=4$, $J_{2}/J_{1}=0.5$. Selected as a high magnetic frustration
point, with weaker transverse spin interactions softening quantum
fluctuations.
\end{itemize}
We found important qualitative differences between the first case
(Case 1 in the following). and the other three. For this reason we
provide details on that and the second one (Case 2 in the following)
and defer the others for Supplemental Material {[}\onlinecite{SM}{]}.
Our complete analysis leads to the schematic phase diagram shown in
Figure \ref{fig: phase-diagram}.

We have checked at each of these points that the magnetization curves
indeed show plateaus at $M=1/3$, with different widths. For Cases
1 and 2 these are shown in Figure \ref{fig: plateau curves}. 
\begin{center}
\begin{figure}
\begin{centering}
\includegraphics[scale=0.55]{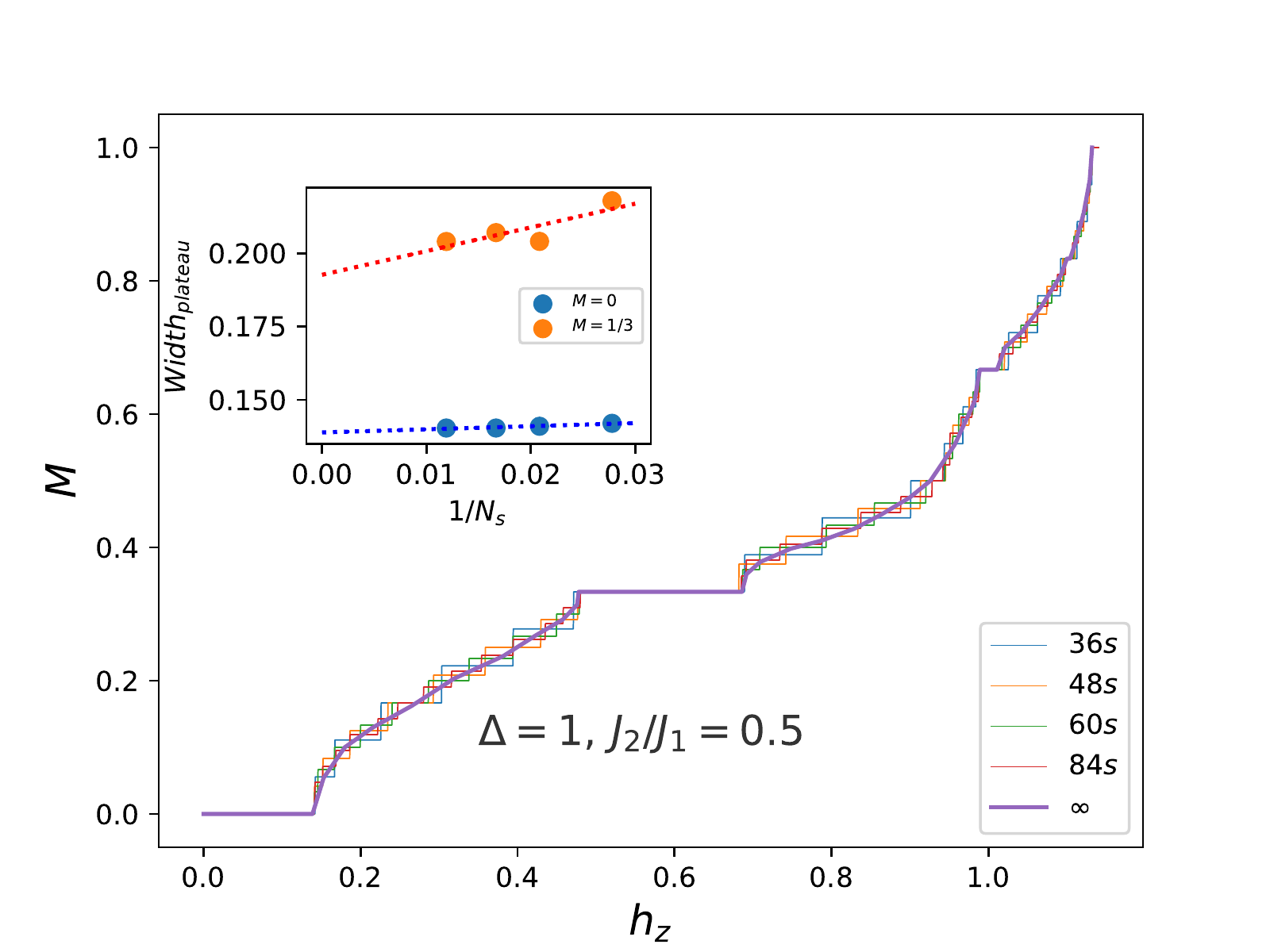}
\par\end{centering}
\begin{centering}
\includegraphics[scale=0.55]{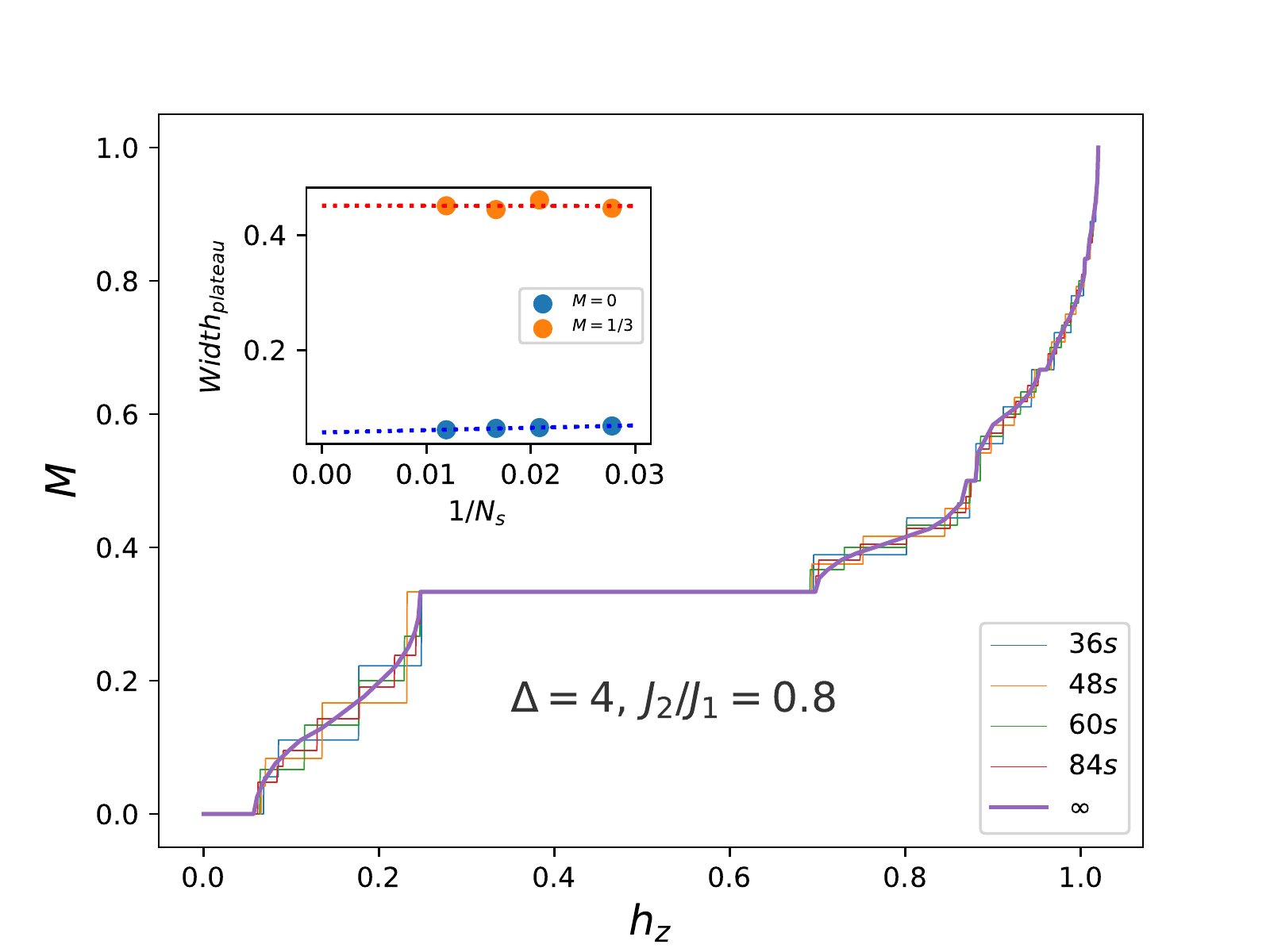}
\par\end{centering}
\caption{\label{fig: plateau curves} Magnetization curves showing the existence
of the $M=1/3$ plateau in Case 1 (upper plot) and Case 2 (lower plot)
discussed in detail in this work. The widths of the main magnetization
plateaus are computed for several finite chain lengths and extrapolated
to infinite size as shown in the insets. The smooth lines suggest
the magnetization curves in the thermodynamic limit. For the Case
1 the plateau is narrower than the Case 2, due to the enhancement
of quantum fluctuations favored by isotropic interactions and high
magnetic frustration.}
\end{figure}
\par\end{center}

\medskip{}

\textbf{Case 1} ($\Delta=1$, $J_{2}/J_{1}=0.5$): we have found that
the \emph{double frustration} caused by the dipolar degrees of freedom
is able to radically change the otherwise classical magnetic plateau
structure to a quantum one. This is a manifestation of a \emph{strong
magneto-electric effect.} The local profile of the relevant variables
has been computed in a chain of $N_{s}=174$ sites, with periodic
boundary conditions and a magnetic field $h_{z}=0.6$ setting $M=1/3$
(meaning $S_{total}^{z}=\left(N_{s}\cdot\frac{1}{2}\right)/3$). 

The local results show a repeated structure every three sites, as
expected; a detail of a portion of the chain is shown in the upper
panel of Figure \ref{fig: quantum plateau}. We have used three color
tones (light-medium-dark) to identify the corresponding period three
sub-lattices. We have also drawn vertical lines in these plots to
indicate the magnetic sites, drawing site variables ($\langle S_{i}^{z}\rangle$)
markers upon these lines and bond variables ($\delta_{i}$, $p_{i}$
and spin-spin correlations) markers between them. 

One can see in the sequence of $\langle S_{i}^{z}\rangle$ (red circles
in the upper plot, with light-medium-dark tones every three sites)
a repetition of one spin up ($\langle S^{z}\rangle\approx0.4$, in
medium red) followed by two spins with almost vanishing expectation
value ($\langle S^{z}\rangle\approx0$, in dark and light red). The
nearest neighbors spin correlations (red diamonds in the lower plot,
with corresponding light-medium-dark tones every three bonds) take
a very negative value (below $-0.6$) every three bonds, indicating
the tendency to form two sites local quantum singlets just between
sites with almost vanishing $\langle S^{z}\rangle$, with low antiferromagnetic
correlations between them and sites with spin up. The longitudinal
and transverse correlations are shown with up-triangles and horizontal-triangles
for more detail: typical singlet correlations (in dark red) get equal
contributions from each spin component $\langle S_{i}^{x}S_{i+1}^{x}\rangle=\langle S_{i}^{y}S_{i+1}^{y}\rangle=\langle S_{i}^{z}S_{i+1}^{z}\rangle\approx-0.2$,
while the other bonds show almost uncorrelated $z$-components $\langle S_{i}^{z}S_{i+1}^{z}\rangle\approx0$.
The elastic distortions (blue squares in the upper plot, also with
light-medium-dark tones) are negative in the dark bonds and positive
in the rest forming a \textquotedbl short-long-long\textquotedbl{}
(S-L-L) bond distortion pattern. The magnetic ions in the spin quantum
singlets get closer, augmenting the spin exchange $J_{1}\left(1-\alpha\delta_{i}\right)$
for better magnetic energy gain at the expense of elastic energy cost.
These together are clear signals of the $\bullet\!\!-\!\!\bullet\uparrow$
quantum plateau structure (see Figure \ref{fig: simple patterns}-B,
where the same tones of red are used for spin sites). The $\Uparrow\Uparrow\Downarrow$
dipole amplitudes (green diamonds in the upper plot) pin the dipoles
pointing down in the short bonds; this makes antiparallel dipoles
get closer and parallel dipoles get farther, in a pattern that minimizes
the electro-elastic energy (see Figure \ref{fig: simple patterns}-C,
where the same tones of green are used for dipoles). The same elastic
distortions thus contribute to the gain of both electric and magnetic
energy. Notice that this ground state breaks the translation invariance
of the Hamiltonian in Eq. (\ref{eq: H-full-expanded}) but maintains
the inversion symmetry with respect to spin up sites or dipole down
bonds (in contrast to Case 2 discussed below). In consequence the
ground state is three-fold degenerate.

One should recall that a different (classical) magnetic order has
been previously observed at the $M=1/3$ plateau of the isotropic,
frustrated $J_{1}-J_{2}$ antiferromagnetic-elastic spin $S=1/2$
chain {[}\onlinecite{2007-Gazza-etal}{]} in the absence of local
dipoles. We can say that the distortions associated to the dipolar
order dominate and destroy the otherwise collinear $\uparrow\uparrow\downarrow$
classical magnetic plateau order of the isotropic, frustrated $J_{1}-J_{2}$
antiferromagnetic-elastic spin chain. Instead they give rise to a
$\bullet\!\!-\!\!\bullet\uparrow$ quantum magnetic plateau order,
elastically compatible with the dipolar order, where the formation
of spin quantum singlets lowers the magnetic energy. This is one of
the main results in this work. 

The ground state obtained in the Case 1 may be visually summarized
in the cartoon description provided at the bottom of Figure \ref{fig: quantum plateau}. 
\begin{center}
\begin{figure}
\begin{centering}
\includegraphics[scale=0.5]{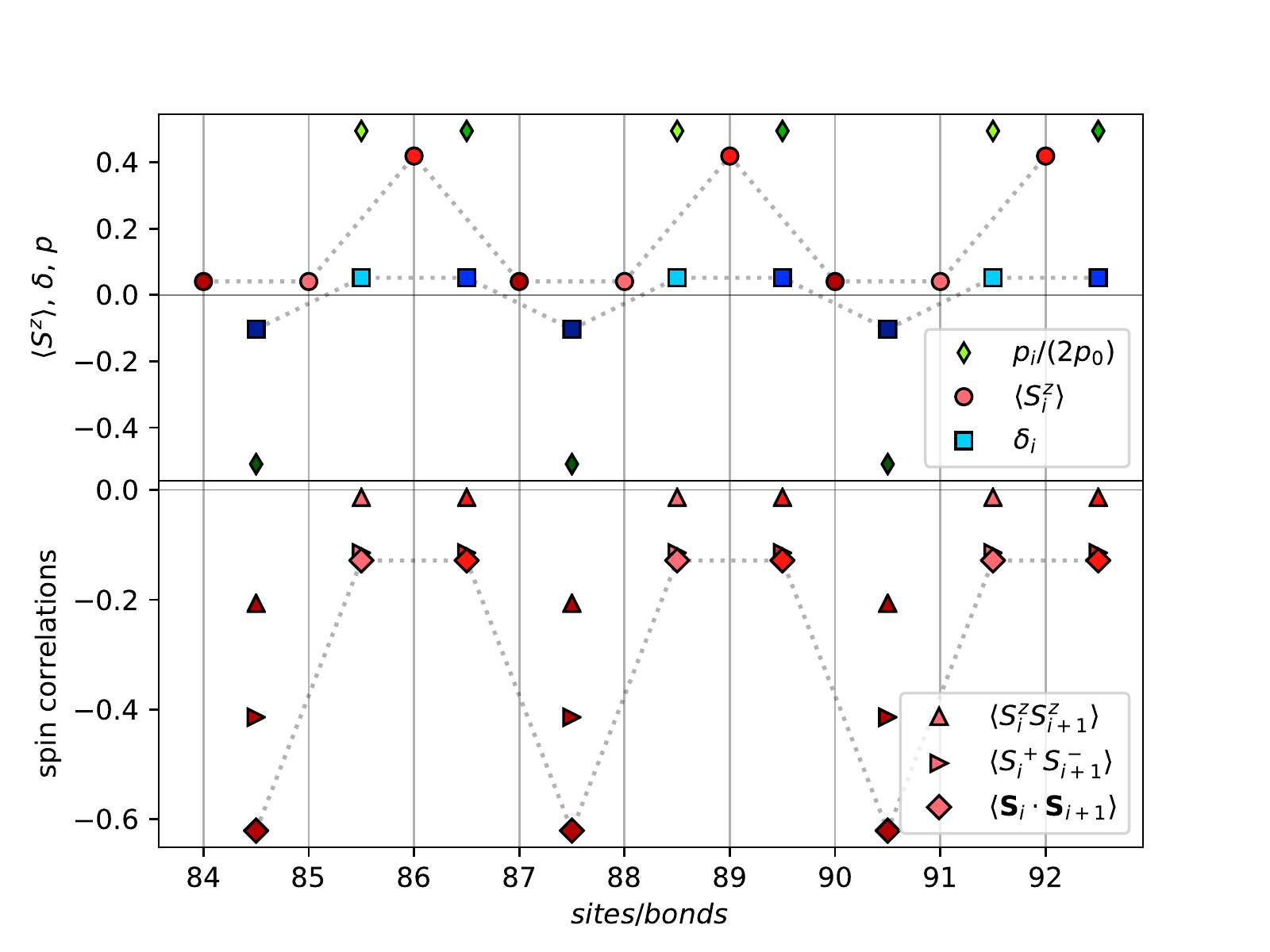}
\par\end{centering}
\begin{centering}
\includegraphics[scale=0.55]{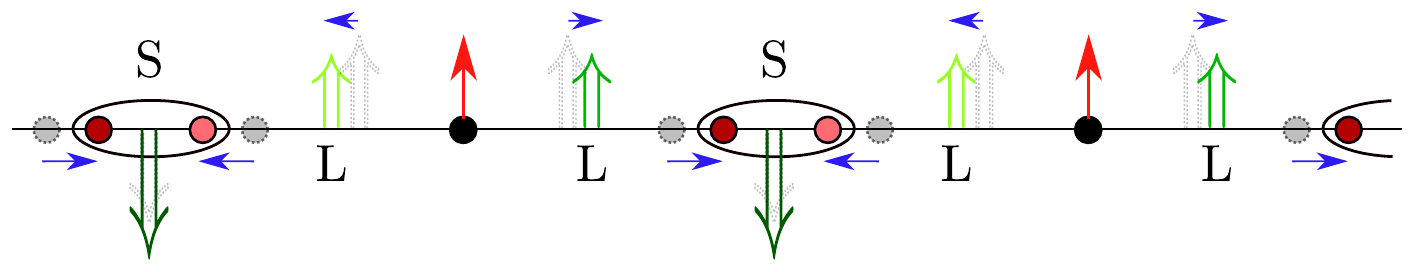}
\par\end{centering}
\caption{\label{fig: quantum plateau} The $M=1/3$ ground state for $\Delta=1$,
$J_{2}/J_{1}=0.5$ (Case 1) obtained with the self-consistent numerical
computations. Different light-medium-dark tones are used to emphasize
the period three structure. The upper plot shows the local profile
of the site observables (spin expectation values $\langle S_{i}^{z}\rangle$
in red circles, vertical lines indicating sites) and bond observables
(distortions $\delta_{i}$ in blue squares and dipolar moments $p_{i}$
(divided by $2p_{0}$) in green thin-diamonds, drawn between sites).
The lower plot shows the nearest neighbors spin correlations $\langle\boldsymbol{S}_{i}\cdot\boldsymbol{S}_{i+1}\rangle$
in red diamonds; a detail of longitudinal $\langle S_{i}^{z}S_{i+1}^{z}\rangle$
and transverse $\langle S_{i}^{+}S_{i+1}^{-}\rangle=\langle S_{i}^{x}S_{i+1}^{x}\rangle+\langle S_{i}^{y}S_{i+1}^{y}\rangle$
correlations is given in triangles. Data shows the formation of local
quantum singlets alternating with partially decoupled spins up every
three sites. The cartoon picture at the bottom qualitatively collects
these numerical results and shows the compatibility of the \textquotedbl short-long-long\textquotedbl{}
(S-L-L) quantum magneto-elastic and electro-elastic patterns in Figure
\ref{fig: simple patterns}.}
\end{figure}
\par\end{center}

\medskip{}

\textbf{Case 2} ($\Delta=4$, $J_{2}/J_{1}=0.8$): in the anisotropic,
less frustrated case we have observed qualitatively different magnetic
and electric orders, again with a period three structure. Numerical
results are shown in Figure \ref{fig: anisotropic frustrated classical parity broken plateau}
together with the corresponding cartoon picture. 

The spins clearly adopt the $\uparrow\uparrow\downarrow$ classical
plateau structure. This is seen in the sequence of $\langle S_{i}^{z}\rangle$
with two positive, one negative values close to $0.5$ (red circles
in the upper plot) and mainly in the almost vanishing transverse correlations
$\langle S_{i}^{+}S_{i+1}^{-}\rangle$ (horizontal-triangles in the
lower plot); longitudinal correlations close to $0.25\,(-0.25)$ (up-triangles
in the lower plot) correspond to collinear parallel (antiparallel)
spins. In this state the magnetic sector could be well described by
classical Ising spins, neglecting the quantum fluctuations. However,
we find in the next section that the magnetic excitations above this
plateau state show a clear quantum behaviour.

The novelty here is that the lattice distortions do not follow the
pattern of the magnetic correlations (compare with Figure \ref{fig: simple patterns}-A).
The first neighbor dipolar terms in the self-consistent Eq. (\ref{eq: self consistence}),
following the $\Uparrow\Uparrow\Downarrow$ configuration induced
by the external electric field (see Figure \ref{fig: simple patterns}-C),
are not compatible with such magnetic correlations and force a competition
in determining the lattice distortions. The resulting distortion pattern
does neither optimize the magnetic energy nor the dipolar energy separately,
but their sum with the elastic energy. It can be qualitative described
as \textquotedbl long-null-short\textquotedbl{} (L-0-S) distortion
pattern (following light-medium-dark blue squares in the upper plot
of Figure \ref{fig: anisotropic frustrated classical parity broken plateau}).
A different, degenerate, ground state is obtained by inversion with
respect to any of the \textquotedbl short\textquotedbl{} bonds. Thus
the ground state is six-fold degenerate.
\begin{center}
\begin{figure}[H]
\begin{centering}
\includegraphics[scale=0.5]{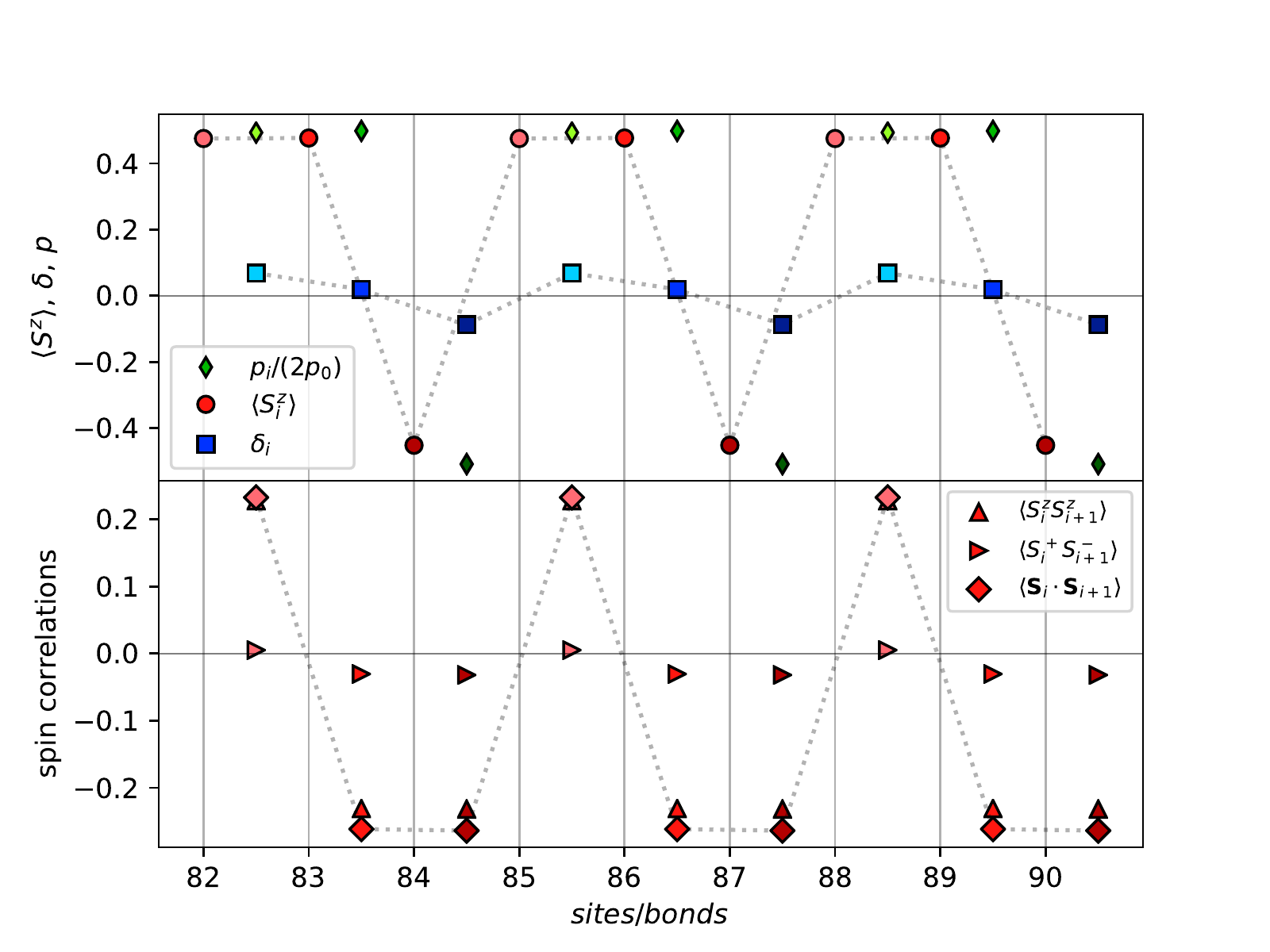}
\par\end{centering}
\begin{centering}
\includegraphics[scale=0.55]{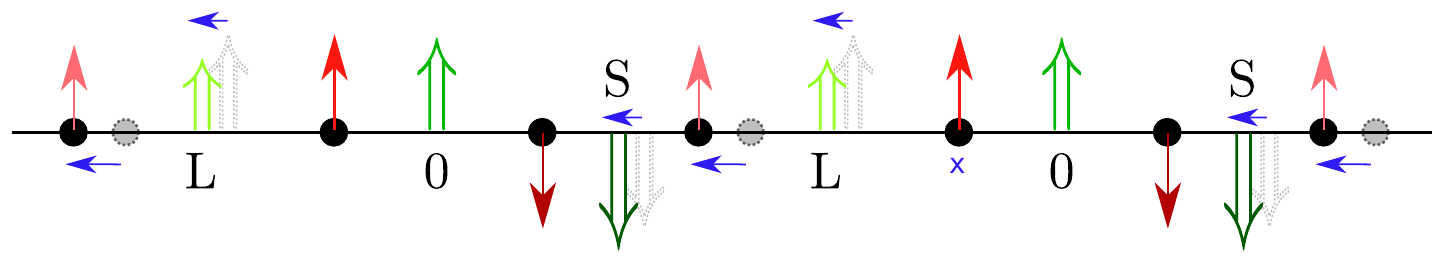}
\par\end{centering}
\caption{\label{fig: anisotropic frustrated classical parity broken plateau}
The $M=1/3$ ground state for $\Delta=4$, $J_{2}/J_{1}=0.8$ (Case
2). Symbols and colors follow the conventions in Figure \ref{fig: quantum plateau}.
Spin data shows the classical $\uparrow\uparrow\downarrow$ plateau
features, with $\langle S_{i}^{z}\rangle$ not reaching $\pm0.5$
because of small quantum fluctuations. However, the elastic distortions
do not fit the classical magneto-elastic pattern in Figure \ref{fig: simple patterns}
but the \textquotedbl long-null-short\textquotedbl{} (L-0-S) sequence
depicted qualitatively in the bottom panel. The interaction between
dipoles and spins, mediated by distortions, frustrates the magneto-elastic
order producing the breaking of inversion symmetry.}
\end{figure}
\par\end{center}

\medskip{}

The features of the ground state in Case 2 may be recognized in the
cartoon in the lower panel of Figure \ref{fig: anisotropic frustrated classical parity broken plateau}.
The double frustration effect leads to a compromising distortion pattern
that breaks the inversion symmetry. This lack of inversion symmetry
in the distortion pattern of course modifies the magnetic and dipolar
couplings through the magneto-elastic coupling $\alpha$ and the pantograph
electro-elastic coupling $\beta$. In consequence the spins (dipoles)
pointing up have slightly different values of $\langle S^{z}\rangle$
($p_{i}$), breaking the inversion symmetry observed in the quantum
plateau Case 1 and also in the $J_{1}-J_{2}$ magneto-elastic spin
chain {[}\onlinecite{2007-Gazza-etal}{]}. This spontaneous symmetry
breaking is another important result, consequence of the present double
frustration effect. Were charge degrees of freedom included, the induced
charge order would result in a longitudinal component of electrical
polarization {[}\onlinecite{2008-vdBrink-Khomskii}{]}.

\medskip{}

The analysis of the other points indicated in Figure \ref{fig: phase-diagram}
show a classical plateau behaviour similar to that observed in Case
2. They correspond to isotropic spin interactions and low frustration
($\Delta=1$, $J_{2}/J_{1}=0.8$) and to a highly frustrated case
with important easy-axis anisotropy ($\Delta=4$, $J_{2}/J_{1}=0.5$).
It appears that both isotropy and high frustration are necessary to
stabilize the quantum plateau. Further numerical exploration indicates
that there is a finite small region around $\Delta=1$, $J_{2}/J_{1}=0.5$
where the $M=1/3$ plateau remains open and the spins order in the
quantum $\bullet\!\!-\!\!\bullet\uparrow$ structure, as shown schematically
in Figure \ref{fig: phase-diagram}. 

\medskip{}

As a summary of this Section, we have provided a qualitative description
and numerical evidence for a novel double frustration effect in a
multiferroic model scenario. 

\section{Composite excitations induced by a magnetic field\label{sec:Composite-excitations}}

The $\Delta S^{z}=1$ spin excitations induced by a magnetic field
on a plateau state are the key to understand the high field plateau
border. It is well known that the excitation of $M=0$ plateau in
one dimensional antiferromagnetic spin chains is not a stable singlet-triplet
excitation but decays into two spinons {[}\onlinecite{1980-Nakano-Fukuyama,1981-Fadeev-Takhtajan,1996-Kiryukhin,1997-Feiguin,2001-Dobry}{]}.
Each spinon carries spin $S^{z}=1/2$ as a topological charge and
may be described as a soliton quasiparticle interpolating between
different dimerized vacua; spatially, the soliton profile can be seen
as a smooth domain wall. We have discussed this spin fractionalization
phenomenom in the present multiferroic model in Ref. I.

The $\Delta S^{z}=1$ spin excitations on top of the $M=1/3$ magnetization
plateau in antiferromagnetic magneto-elastic spin chains is also known
to exhibit spin fractionalization {[}\onlinecite{2007-Gazza-etal}{]}.
Remarkably, this goes beyond the spinon description: the excitation
decays into three $S^{z}=1/3$ noninteracting solitonic excitations
(dubbed tertions). For a classical $\uparrow\uparrow\downarrow$ plateau,
it has been shown that the tertions have a local singlet core causing
the $\uparrow\uparrow\downarrow$ order on one side to be shifted
by one site with respect to the other side. In this way the tertion
interpolates between two different $\uparrow\uparrow\downarrow$ domains. 

It has been observed in several systems that the high field plateau
border is characterized by a sudden finite magnetization jump when
the magnetic field takes a threshold value. The magnetization curves
in Figure \ref{fig: plateau curves} suggest that this might also
occur in our model. If this is the case, the magnetic excitation would
fractionalize into a periodic lattice of self avoiding solitons {[}\onlinecite{1996-Kiryukhin,1998-Lorenz}{]}.
Such a periodic magnetic structure could be detected by unusual line
shapes in neutron scattering data {[}\onlinecite{1999-Horvatic}{]},
as well as the associated lattice distortions could be detected by
X-ray measurements.

A natural question arises, whether these features are modified by
the double frustration effect in the present magneto-electro-elastic
chain. We have explored the numerical self-consistent solutions of
Eqs. (\ref{eq: self consistence}, \ref{eq: constraint}) in periodic
chains with $N_{s}$ sites in the subspace of $S_{total}^{z}=$$\left(N_{s}\cdot\frac{1}{2}\right)/3+1$
(that is one unit of magnetization above $M=1/3$). We considered
large chains in order to allow for a most clear spatial separation
of the three expected tertions \footnote{For a periodic pattern with magnetization $S_{total}^{z}=$$\left(N_{s}\cdot\frac{1}{2}\right)/3+1$
to be commensurate with the chain length, $N_{s}$ must be restricted
to integers of the form $9q+3$ {[}\onlinecite{2007-Gazza-etal}{]}.}. We report results on chains of $N_{s}=174$ sites, where the plateau
state has $S_{total}^{z}=29$ and the excited state has $S_{total}^{z}=30$.

In the present multiferroic model scenario we have confirmed that
the $\Delta S^{z}=1$ excitation induced by a magnetic field on top
of the $M=1/3$ state indeed fractionalizes into three $S^{z}=1/3$
spatially separated tertions. The trial of different dipolar configurations
has shown that the dipolar sector suffers a spontaneous unit polarization
change along the electric field direction to minimize the energy cost
of the distortions accompanying the magnetic order. This polarization
change induced by a magnetic field is an emergent magneto-electric
effect mediated by elastic distortions. We discuss below the numerical
data supporting these statements. 

\medskip{}

\textbf{Case 1}: We show in Figure \ref{fig: quantum plateau excitation}
numerical results for the magnetically excited state in the isotropic
frustrated regime ($\Delta=1$, $J_{2}/J_{1}=0.5$). Using the same
color codes as in the plateau state (see Figure \ref{fig: quantum plateau}),
in the upper panel we show the local $\langle S_{i}^{z}\rangle$ in
red circles, the dipoles $p_{i}$ in green diamonds and the distortions
$\delta_{i}$ in blue squares; the light-medium-dark tones for sites
1, 2, 3 are repeated every three sites to visually distinguish the
associated sub-lattices. In the lower panel we show the $\langle\boldsymbol{S}_{i}\cdot\boldsymbol{S}_{i+1}\rangle$
correlations. With the help of the color tones one can see a short
wavelength oscillation of each observable, with period three as in
the plateau state, but modulated by a long wavelength oscillation
spanning three whole periods along the chain. The regions around sites
$\sim26$ (marked as $I$ in the plots), $\sim26+N_{s}/3=\sim84$
(marked as $II$ in the plots), and $\sim26+2\times N_{s}/3=\sim142$
show locally the same features as the plateau state. However, in the
first one the spin up is located on the medium-red sub-lattice (region
$I$), in the second one it falls on the dark-red sub-lattice (region
$II$) and in the third region it corresponds to the light-red sub-lattice.
Thus, regarding the spin sector, each of these regions adopts one
of the three degenerate possible quantum plateau configurations, different
because of a relative shift of the spin up and the spin singlet positions
by one site to the right. In the sites between the plateau regions
one can see a smooth sub-lattice interpolation between observables;
for instance, the $\langle S_{i}^{z}\rangle$ in the medium-red sub-lattice
evolves from a spin up in region $I$ to a spin in a singlet in region
$II$. These intermediate sites then allocate the solitonic excitations
interpolating between different vacua (in the sense of degenerate
plateau states related by translations); the analysis of local $\langle S_{i}^{z}\rangle$
values shows that they carry a fraction $S^{z}=1/3$ of the magnetic
excitation. At the center of the soliton the spins take a $\downarrow\uparrow\uparrow\downarrow$
configuration, that may be called a classical core between quantum
orders. One can of course notice that the soliton regions occupy an
important fraction of the chain length. As the spatial width of the
magnetic solitons is usually in inverse relation with the plateau
width (or the spin gap producing it {[}\onlinecite{1997-Feiguin}{]}),
we expect that in a larger chain they will maintain their size and
more space will be left for better defined plateau regions.
\begin{center}
\begin{figure}[H]
\begin{centering}
\includegraphics[scale=0.58]{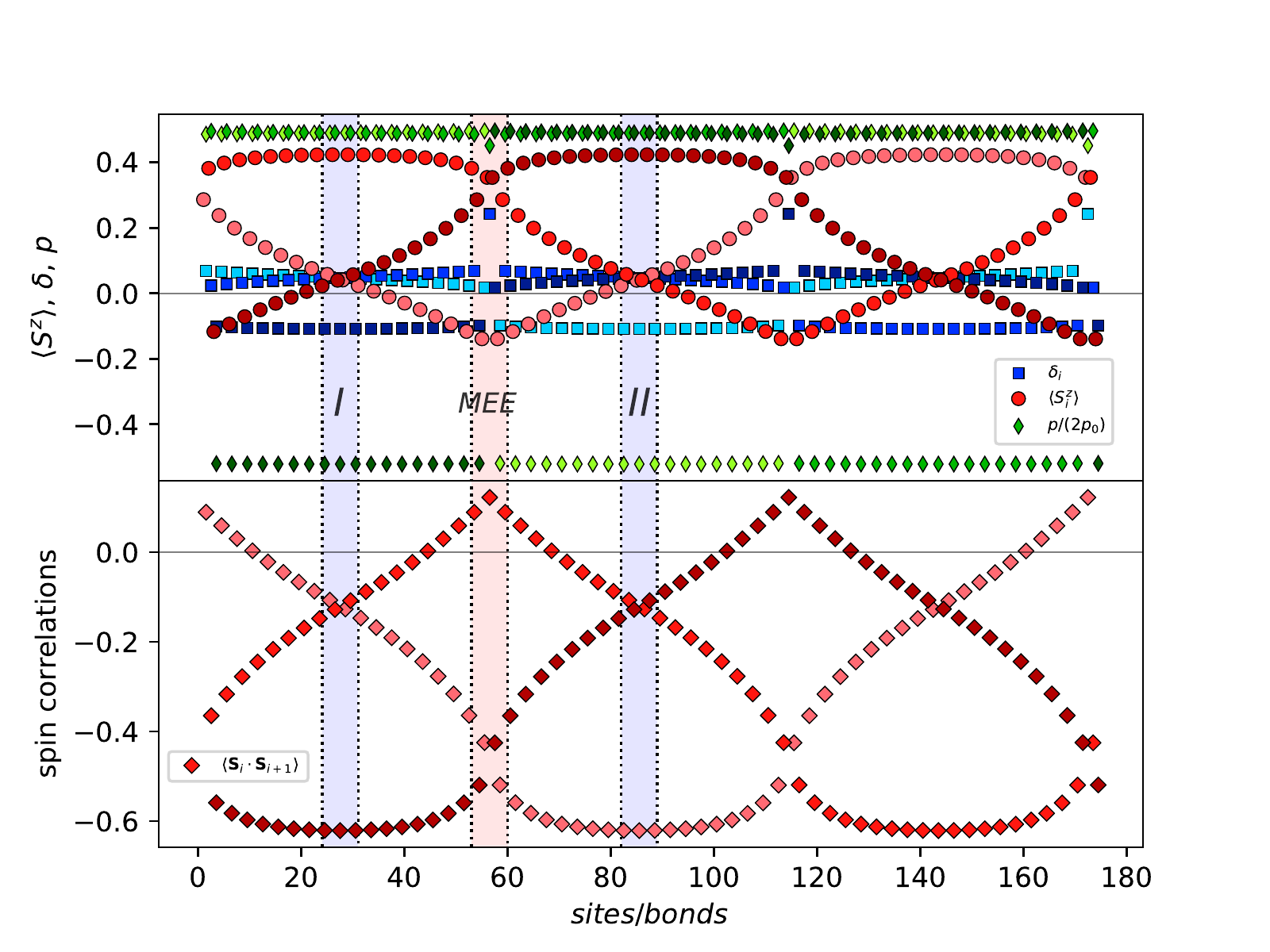}
\par\end{centering}
\caption{\label{fig: quantum plateau excitation} Magneto-electro-elastic excitations
(\emph{MEE}) with $\Delta S^{z}=1$ induced by the magnetic field
on top of the $M=1/3$ plateau for $J_{2}/J_{1}=0.5$, $\Delta=1$
(Case 1). In the top panel we show numerical results for the local
observables (upper plot) and spin correlations (lower plot). Symbols
and colors follow the conventions in Figure \ref{fig: quantum plateau}.
The magnetic sector shows regions which adopt the $\bullet\!\!-\!\!\bullet\uparrow$
quantum plateau order (two of them highlighted as $I$ and $II$),
interpolated by soliton structures with a classical $\downarrow\uparrow\uparrow\downarrow$
center. The position of the $\bullet\!\!-\!\!\bullet\uparrow$ pattern
is shifted to a different sub-lattice across each soliton. Dipolar
domain walls $\Downarrow\Uparrow\Uparrow\Uparrow\Downarrow$ are formed
at the core of the magnetic solitons, shifting the $\Uparrow\Uparrow\Downarrow$
order from one to another sub-lattice to accompany the magnetic configuration;
these dipolar domain walls carry fractionalized spontaneous electric
excitations. The lattice distortions also accompany the magnetic configuration
but show singular localized excitations at each dipolar domain wall.
Each region between $\bullet\!\!-\!\!\bullet\uparrow$ quantum plateau
orders thus allocates a coupled magneto-electro-elastic excitation
(the one between regions $I$ and $II$ is highlighted as \emph{MEE}
in the plots). A cartoon picture of these results is shown in Figure
\ref{fig: quantum plateau excitation-1}.}
\end{figure}
\par\end{center}

The analysis of dipolar configurations necessary to minimize the system
energy has shown that the dipolar order $\Uparrow\Uparrow\Downarrow$
induced by the electric field (fixed at the plateau value $\epsilon=0.16$)
is altered by the appearance of three dipolar domain walls with patterns
$\Downarrow\Uparrow\Uparrow\Uparrow\Downarrow$. Each of them can
be seen as the insertion of an extra $\Uparrow$ dipole, namely a
dipolar excitation. As the three domain walls accumulate one dipole
flip with respect to the homogeneous $\Uparrow\Uparrow\Downarrow$
order, we observe a spontaneous unit dipolar excitation that appears
to decay into three domain walls. A similar behavior is reported in
{[}\onlinecite{2003-Okunishi-b}{]} for magnetic excitations above
the $M=1/3$ plateau in the Ising limit. These dipolar excitations
are localized at the soliton cores, so that the dipolar order observed
at the plateau state is shifted by one site at each domain wall accompanying
the shift of magnetic plateau structures. Notice that magnetic tertions
and dipolar domain walls occur in the same positions, a fact that
may be interpreted as a magneto-electric coupling between magnetic
and dipolar excitations.

The elastic sector evolves smoothly along the magnetic solitons, shifting
the S-L-L pattern by one site as the spins and dipoles do, with a
noticeably exception at the dipolar domain walls. A singularly long
bond is formed there, while the others abruptly interchange from \textquotedbl null\textquotedbl{}
to \textquotedbl short\textquotedbl . We interpret this feature
as a local elastic excitation, coupled to the magneto-electric one.
Thus a localized magneto-electro-elastic excitation (indicated as
\emph{MEE} in Figure \ref{fig: quantum plateau}) shows up between
quantum plateau regions.

A cartoon picture of these features is drawn in Figure \ref{fig: quantum plateau excitation-1}.
The spins, dipoles and distortions are schematically indicated at
the quantum plateau regions $I$ and $II$ (as labeled in Figure \ref{fig: quantum plateau excitation}),
separated by the \emph{MEE} excitation with a classical magnetic $\downarrow\uparrow\uparrow\downarrow$
core coinciding with the $\Downarrow\Uparrow\Uparrow\Uparrow\Downarrow$
dipolar domain wall.

\begin{widetext} 
\begin{center}
\begin{figure}[H]
\begin{centering}
\includegraphics[scale=0.5]{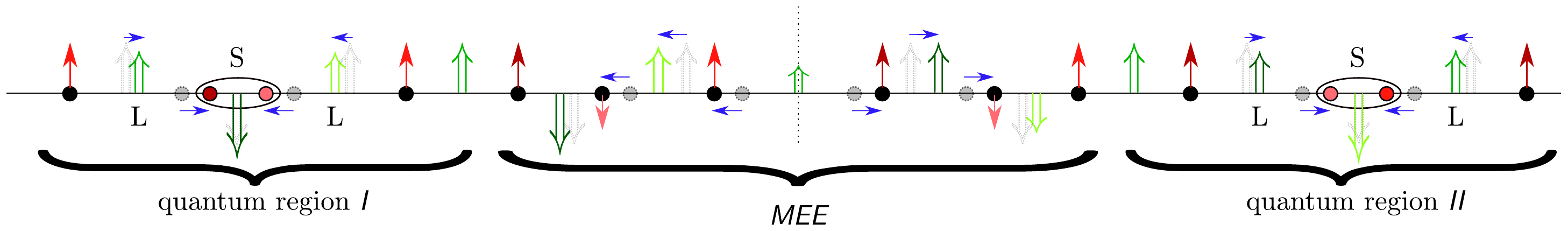}
\par\end{centering}
\caption{\label{fig: quantum plateau excitation-1} Qualitative picture of
the magneto-electro-elastic (\emph{MEE}) excitation separating the
quantum plateau regions \emph{I} and\textsf{ }\emph{II} in Figure
\ref{fig: quantum plateau excitation}. The excitation has a classical
magnetic $\downarrow\uparrow\uparrow\downarrow$ core coinciding with
the $\Downarrow\Uparrow\Uparrow\Uparrow\Downarrow$ dipolar domain
wall and a highly enlarged bond. Local observables and color codes
follow actual data in the upper plot in Figure \ref{fig: quantum plateau excitation}.
The dotted line is a mirror plane showing the parity symmetry of the
\emph{MEE} excitation. The local order in the quantum region \emph{II}
is shifted by one site with respect to the order in the quantum region
\emph{I}.}
\end{figure}
\par\end{center}

\end{widetext}

\medskip{}

\textbf{Case 2}: In Figure \ref{fig: classical anisotropic plateau excitation}
we show the qualitatively different results obtained for the excitations
of the classical plateau state in the anisotropic less frustrated
regime $J_{2}/J_{1}=0.8$, $\Delta=4$. Here the magnetic sector,
in the $S_{total}^{z}=$$\left(N_{s}\cdot\frac{1}{2}\right)/3+1$
excited subspace, presents three regions with each $\uparrow\uparrow\downarrow$
classical plateau order, periodically modulated along the chain. In
each classical plateau region the $\uparrow\uparrow\downarrow$ pattern
lies in different sub-lattices (for instance regions \emph{I} and\textsf{
}\emph{II} in the Figure), separated by solitons (one is highlighted
as a magnetic soliton (\emph{MS}) in the plot). Again these solitons
are tertions, carrying a fraction $S^{z}=1/3$ of the magnetic excitation.
They interpolate classical plateau regions, having a local quantum
spin singlet core; the same features have been observed in the excitations
of the $M=1/3$ classical plateau in magneto-elastic chains, in absence
of dipolar degrees of freedom {[}\onlinecite{2007-Gazza-etal}{]}.
Besides, the solitons are narrower than in Case 1, in accordance with
a wider magnetic plateau (see Figure \ref{fig: plateau curves}).
\begin{center}
\emph{}
\begin{figure}[H]
\begin{centering}
\includegraphics[scale=0.58]{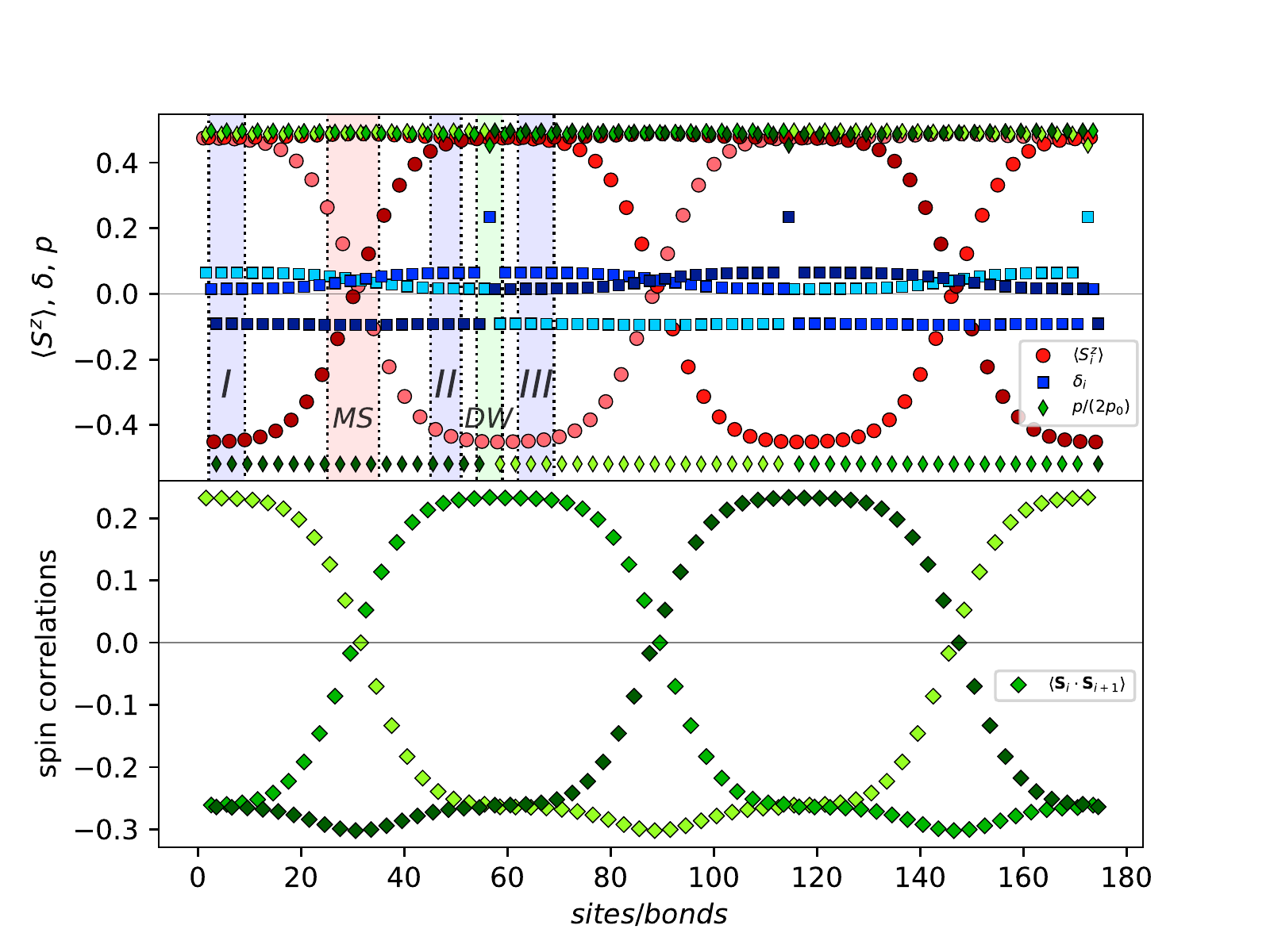}
\par\end{centering}
\caption{\label{fig: classical anisotropic plateau excitation} Decoupled electro-elastic
and magnetic and excitations with $\Delta S^{z}=1$ induced by the
magnetic field on top of the $M=1/3$ plateau for $J_{2}/J_{1}=0.8$,
$\Delta=4$ (Case 2). The plot in the upper panel shows the local
observables, that in the lower panel shows spin correlations. Magnetic
solitons (one of them labeled as\textsf{ }\emph{MS} in the Figure)
having a singlet core interpolate between different classically $\uparrow\uparrow\downarrow$
ordered regions related by one-site translations (see for instance
regions \emph{I} and \emph{II} in the Figure). Dipolar excitations
in the form of domain walls $\Downarrow\Uparrow\Uparrow\Uparrow\Downarrow$
(\emph{DW}), accompanied by localized elastic excitations, are spatially
decoupled from the magnetic solitons. They separate different mirror-related
classically $\uparrow\uparrow\downarrow$ ordered regions (see for
instance \emph{II} and \emph{III}). All of the six degenerate magneto-electro-elastic
classical plateau configurations show up in different regions of the
system. A cartoon picture of these results is shown in Figure \ref{fig: classical anisotropic plateau excitation-1}.}
\end{figure}
\par\end{center}

The dipolar sector again presents a unit spontaneous excitation (dipole
flip) fractionalized into three domain walls (one of them is highlighted
as \emph{DW} in the plot). But in this case the domain walls appear
to decouple from the magnetic solitons. Instead, they occur inside
a $\uparrow\uparrow\downarrow$ plateau region signaling a parity
change of the accompanying lattice distortions (see for instance the
regions highlighted as \emph{II} and \emph{III} in the plot, with
mirror symmetry with respect to the domain wall between them). Singular
elastic excitations (very long bonds) show up together with the dipolar
domain walls. One can thus observe electro-elastic excitations well
decoupled from magnetic $S^{z}=1/3$ excitations. 

As the dipolar domain walls separate the two parity-related degenerate
elastic configurations compatible with the same magnetic order, all
of the six possible (degenerate) magneto-electro-elastic classical
plateau configurations are realized along the system length.

\begin{widetext} 
\begin{center}
\begin{figure}[H]
\begin{centering}
\includegraphics[scale=0.5]{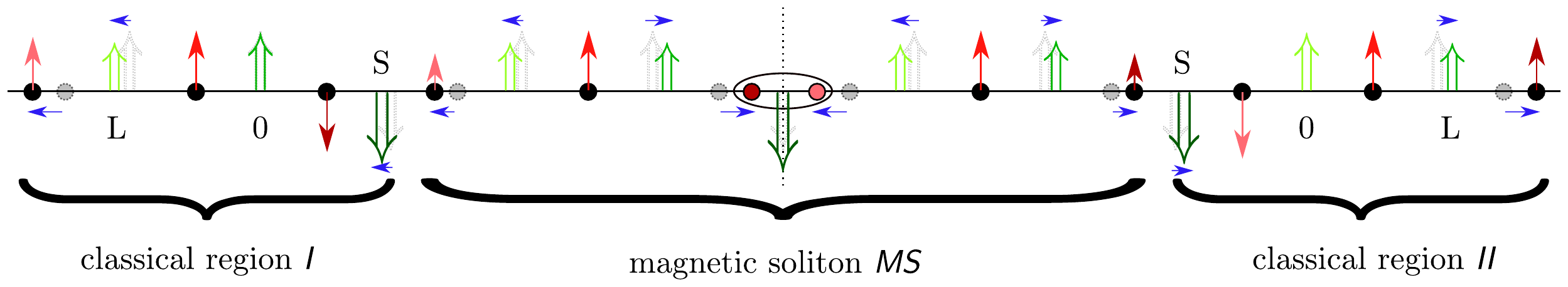}
\par\end{centering}
\begin{centering}
\includegraphics[scale=0.5]{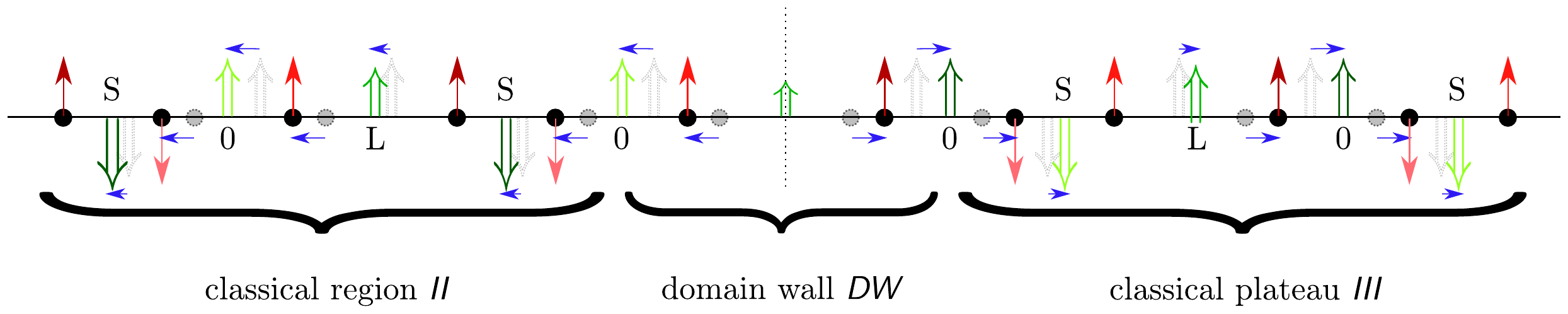}
\par\end{centering}
\caption{\label{fig: classical anisotropic plateau excitation-1} Qualitative
picture of the magneto-elastic and electro-elastic excitations in
Figure \ref{fig: classical anisotropic plateau excitation}. The cartoon
in the upper panel describes the magneto-elastic solitonic excitation
\emph{MS} with a quantum singlet core interpolating the classical
orders highlighted as \emph{I} and \emph{II} in Figure \ref{fig: quantum plateau excitation};
in passing the soliton, the L-0-S distortion pattern changes to S-0-L
and the $\uparrow\uparrow\downarrow$ magnetic pattern is shifted
by one site while the dipolar $\Uparrow\Uparrow\Downarrow$ pattern
remains unaltered. In the lower panel the electro-elastic domain wall
\emph{DW}\textsf{ }separates the classical orders highlighted as \emph{II}
and \emph{III}; the S-0-L distortion pattern changes again to L-0-S
and the dipolar $\Uparrow\Uparrow\Downarrow$ pattern is shifted by
one site but the magnetic $\uparrow\uparrow\downarrow$ order remains
the same at both sides of the domain wall. Dotted lines are added
to make apparent the mirror symmetry of the excitation configurations.}
\end{figure}
\par\end{center}

\end{widetext}

We have drawn a schematic description of these results in Figure \ref{fig: classical anisotropic plateau excitation-1}:
the transition between the classical $\uparrow\uparrow\downarrow$
plateau regions marked as \emph{I} and \emph{II} is given by a magnetic
soliton \emph{MS} passing through a spin singlet, while that between
the classical plateau regions \emph{II} and \emph{III} is given by
an electro-elastic excitation without disruption of the $\uparrow\uparrow\downarrow$
magnetic order.

\medskip{}

It is interesting to recall an argument based on the bosonized description
of the $M=1/3$ plateau in spin $S=1/2$ chains, discussed in {[}\onlinecite{2007-Gazza-etal}{]}.
Along this argument the occurrence of the classical or quantum plateau
are related to the vacuum expectation value of a compactified bosonic
field. This explains why a soliton interpolating classical configurations
must pass over a quantum ordered region as found in that reference
and also here in Case 2. Conversely, the same argument suggests that
a soliton interpolating quantum configurations must pass over a classical
order. As far as we know, our findings in Case 1 are the first realization
of this conjecture.

In comparing the location of dipolar domain walls $\Downarrow\Uparrow\Uparrow\Uparrow\Downarrow$
with respect to the magnetic order, one can extract as a thumb rule
that they fit better in the elastic distortions of a classical $\downarrow\uparrow\uparrow\downarrow$
magnetic environment (see cartoons in Figures \ref{fig: quantum plateau excitation-1}
and \ref{fig: classical anisotropic plateau excitation-1}, second
line). Thus, when exciting a classical plateau state the dipolar domain
walls are located in the classical plateau regions, away from magnetic
excitations. Instead, when exciting a quantum plateau state the dipolar
domain walls are located in the classical core of the magnetic solitons
forming a composite \emph{MEE} quasiparticle.

\section{Summary and conclusions}

We have explored the interplay between frustrated magnetic and dipolar
orders in a one dimensional model for collinear type II multiferroic
materials, where electric and magnetic degrees of freedom are indirectly
coupled by the lattice distortions. More precisely, we have investigated
the commensurability of the $P=1/3$ period three dipolar order $\Uparrow\Uparrow\Downarrow$
with the period three magnetic configurations observed in many frustrated
magnetic materials within $M=1/3$ magnetization plateaus. 

Both from qualitative arguments and extensive DMRG computations we
have found that the dipolar order introduced by frustrating dipolar
interactions competes with the magnetic order set in turn by the magnetic
frustration at the M=1/3 plateau. This opens a non trivial scenario
which we dub \emph{double frustration}. Our analysis provides clear
and surprising effects due to this double frustration. In a regime
of low anisotropy and high magnetic frustration, favoring quantum
fluctuations, the double frustration is responsible for the stabilization
of a \emph{quantum} $M=1/3$ plateau state. In contrast, in all other
cases (either introducing higher easy-axis anisotropy and/or reducing
magnetic frustration) the second frustration competition leads to
the spontaneous parity symmetry breaking in the order of the \emph{classical}
$M=1/3$ plateau state. From this parity breaking mechanism, and in
the presence of charge order along the chain,\textcolor{red}{{} }a longitudinal
component of the polarization should appear {[}\onlinecite{2008-vdBrink-Khomskii}{]}.
Detection of different directions of the polarization could be the
clue to identify the underlying magneto-electric effects operating
in a given material.

We have also discussed the excitations caused by the increase of the
magnetic field. We have found that the $\Delta S^{z}=1$ magnon on
top of the $M=1/3$ state fractionalizes into three $S^{z}=1/3$ spatially
separated solitons, encompassing elastic distortions adapted to the
magnetic order. This change in the distortion pattern induces, in
the dipolar sector, a spontaneous unit polarization change which in
turn fractionalizes into three sharp domain walls. Moreover, on top
of the quantum plateau state these fractional excitations form a composite
magneto-electro-elastic quasiparticle. This emergent magneto-electric
effect, that is the polarization change induced by a magnetic field
mediated by elastic distortions, is one of the main results in the
present paper. 

The nature of the plateau state structure and the appearance of intertwined
magnetic and electric fractional excitations, mediated by the lattice,
are experimentally accessible by neutron scattering for the spin-channel
and by X-ray scattering for the lattice distortions. The striking
differences between the present results and those for pure magneto-elastic
chains are clear signals of the role of dipolar interactions in multiferroic
systems and may guide the search for materials realizing strong magneto-electric
effects. 

\section*{Acknowledgements}

This paper was partially supported by CONICET (Grant No. PIP 2015-813
and No. PIP 2015-364), Argentina.


\begin{thebibliography}{10}
\bibitem{footnote-1}We use double arrows $\Uparrow$, $\Downarrow$
in text and graphics to depict dipole orientations, and single arrows
$\uparrow$, $\downarrow$ to depict spin orientations.
\end{thebibliography}
\end{document}


\title{Supplementary Material for\\
 Double frustration and magneto-electro-elastic excitations in collinear
multiferroic materials}
\author{D. C. Cabra}
\author{A. O. Dobry}
\author{C. J. Gazza }
\author{G. L. Rossini }
\begin{abstract}
We provide here numerical data for the $M=1/3$ magnetic plateau state
and the magnetically excited state in two more parameter regimes,
giving support to the phase diagram of Figure 3 in the main text.
\end{abstract}
\maketitle

\section{Case 3: $J_{2}/J_{1}=0.8$, $\Delta=1$}

These values are chosen as a point with isotropic Heisenberg interaction
and low frustration. We confirm the presence of the $M=1/3$ magnetic
plateau, shown in Fig. \ref{fig: magnetization curve case 3}. The
magnetization and distortion profiles at the plateau state and the
magnetically excited state, shown in Figs. \ref{fig: plateau profile case 3}
and \ref{fig: excited profile case 3}, are very similar to Case 2
in the main text. This is a classical plateau configuration with broken
inversion symmetry, due to the double frustration. Magnetic excitations
in the form of solitons are spatially decoupled from electro-elastic
excitations. 
\begin{center}
\begin{figure}[h]
\begin{centering}
\includegraphics[scale=0.5]{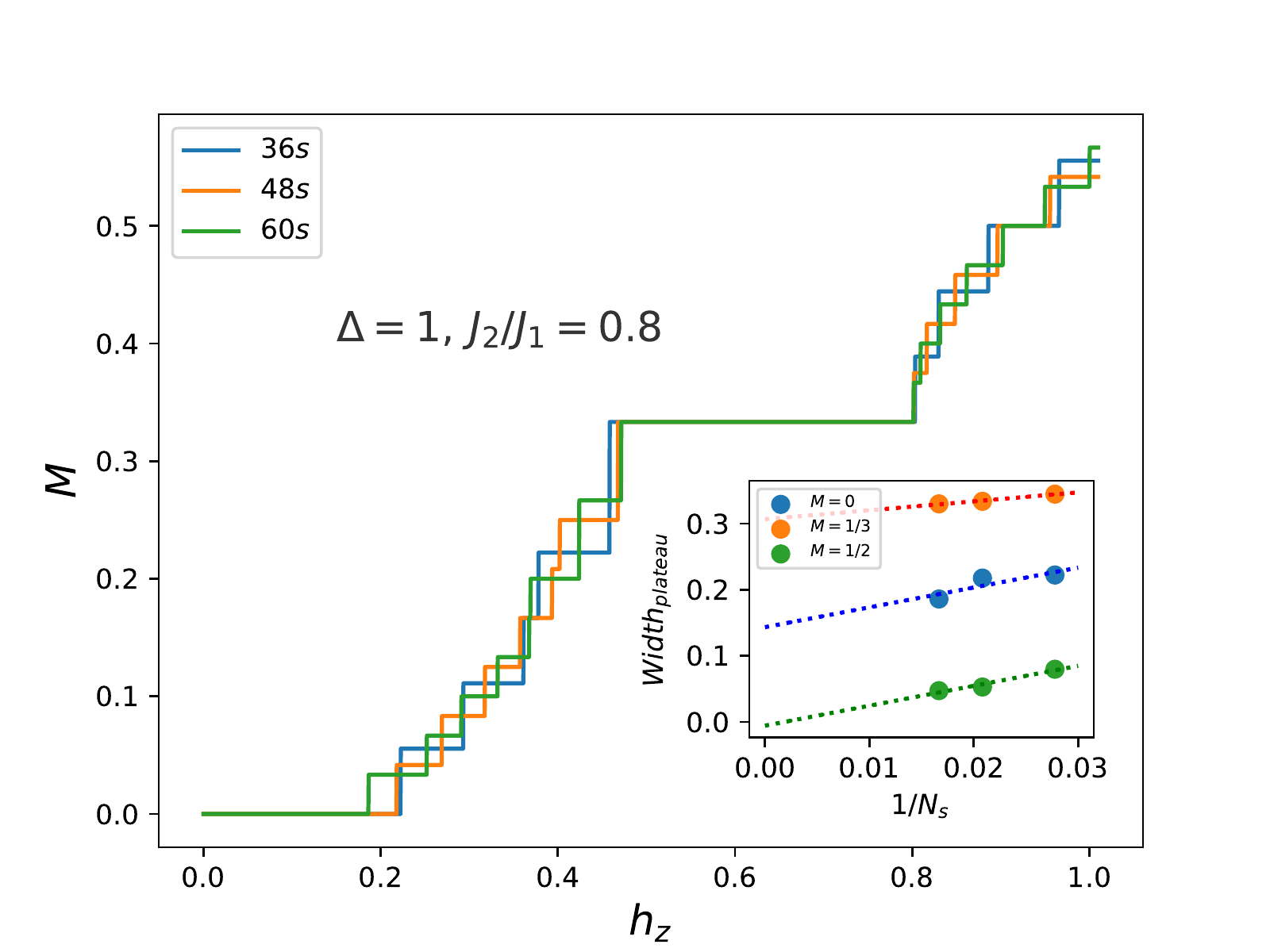}
\par\end{centering}
\caption{\label{fig: magnetization curve case 3} Magnetization curve for Case
3. Size scaling of the plateau widths shown in the inset. }
\end{figure}
\par\end{center}

\begin{center}
\begin{figure}
\begin{centering}
\includegraphics[scale=0.55]{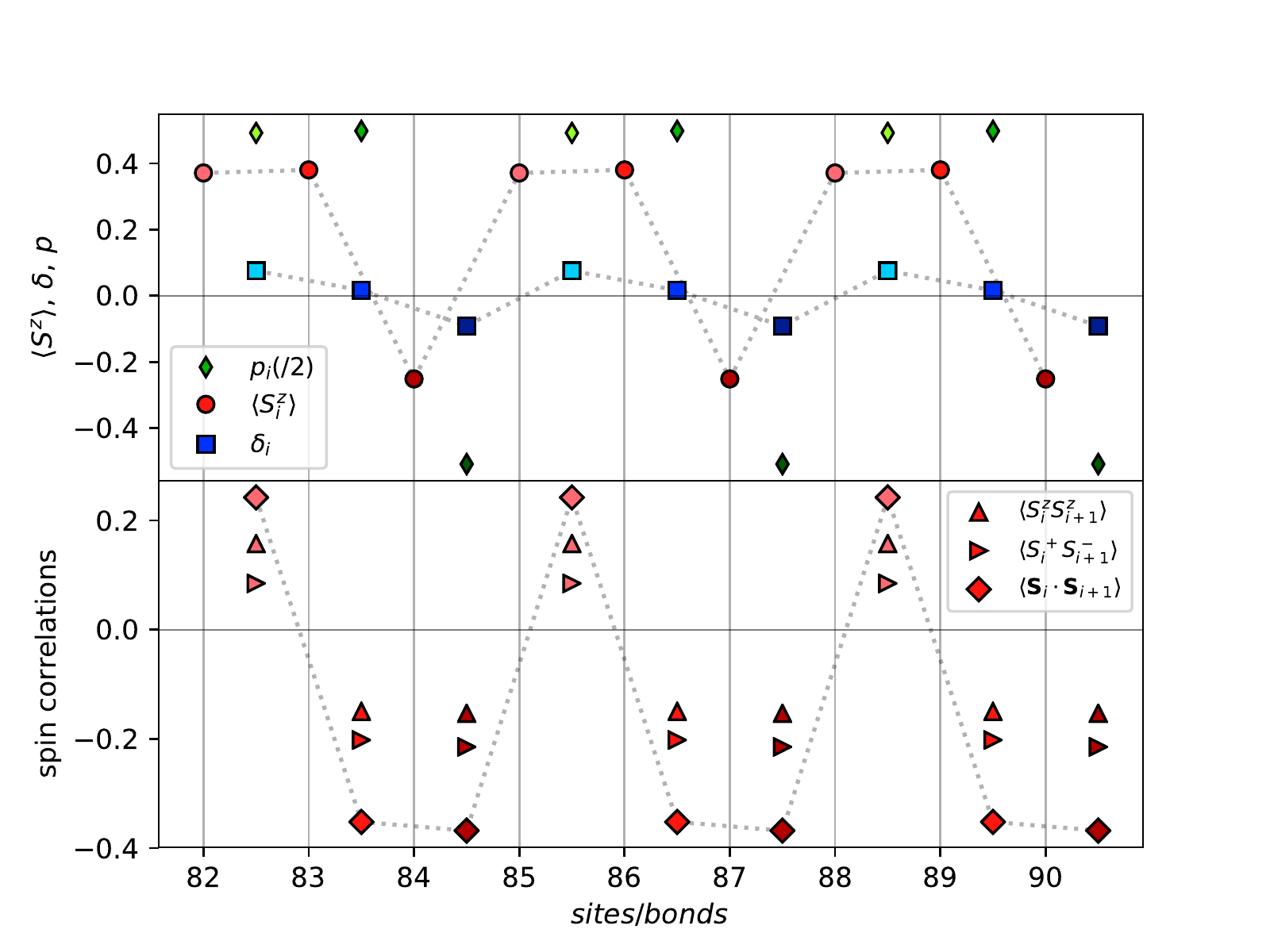}
\par\end{centering}
\caption{\label{fig: plateau profile case 3} $M=1/3$ plateau configuration
for Case 3. Markers as in Fig. 5 in the main text. }
\end{figure}
\par\end{center}

\begin{center}
\begin{figure}[H]
\begin{centering}
\includegraphics[scale=0.55]{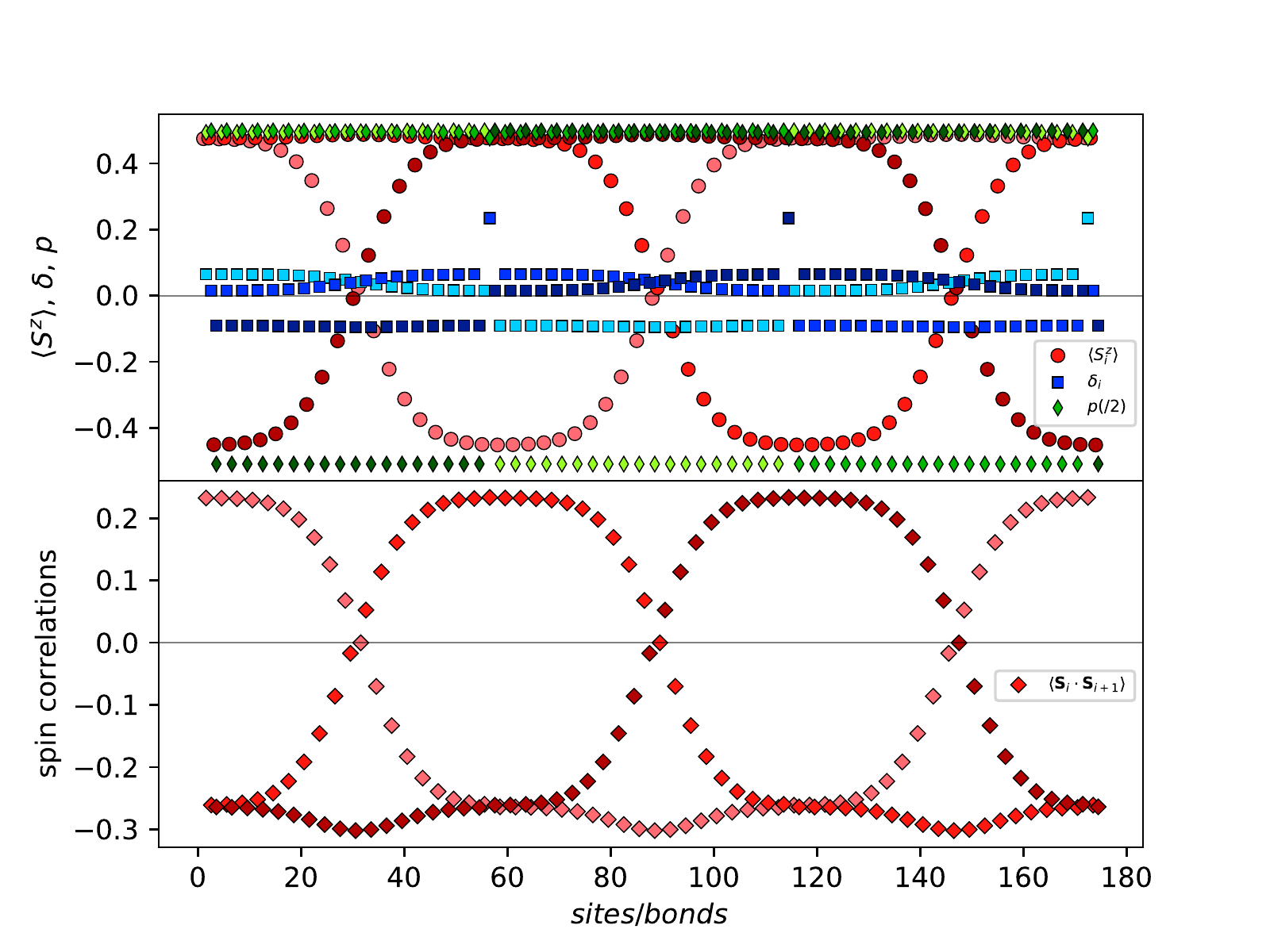}
\par\end{centering}
\caption{\label{fig: excited profile case 3} Magnetically excited state for
Case 3.}
\end{figure}
\par\end{center}

\section{Case 4: $J_{2}/J_{1}=0.5$, $\Delta=4$}

These values are chosen as a highly magnetic frustrated point, with
weaker transverse spin interactions softening quantum fluctuations.
We confirm the presence of the $M=1/3$ plateau, shown in Fig. \ref{fig: magnetization curve case 4}.
The magnetization and distortion profiles at the plateau state and
the magnetically excited state, shown in Figs. \ref{fig: plateau profile case 4}
and \ref{fig: excited profile case 4}, are very similar to Case 2
in the main text and Case 3 above. This is again a classical plateau
configuration with broken inversion symmetry.
\begin{center}
\begin{figure}[h]
\begin{centering}
\includegraphics[scale=0.5]{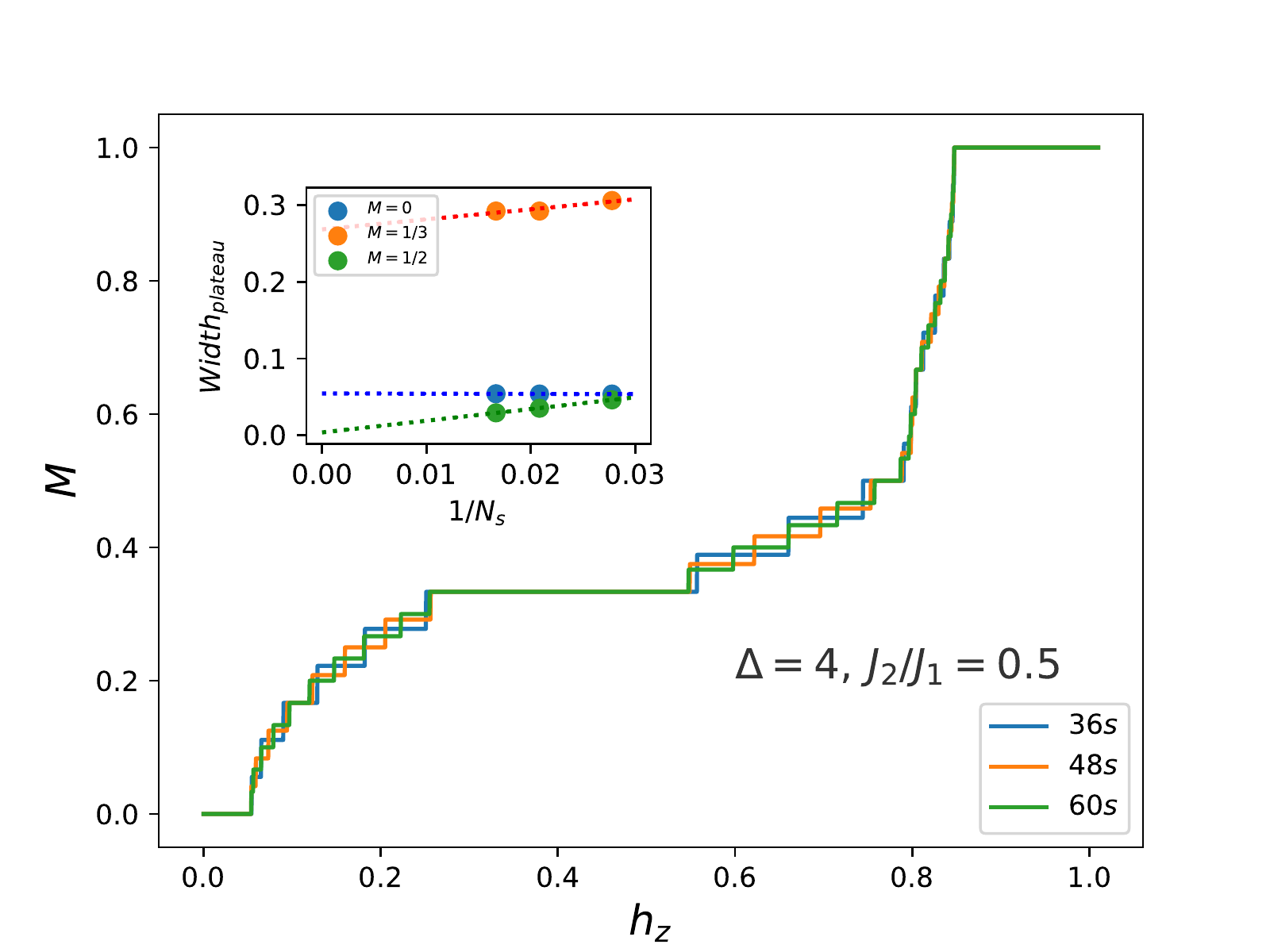}
\par\end{centering}
\caption{\label{fig: magnetization curve case 4}Magnetization curve for Case
4. Size scaling of the plateau widths shown in the inset. }
\end{figure}
\par\end{center}

\begin{center}
\begin{figure}[H]
\begin{centering}
\includegraphics[scale=0.55]{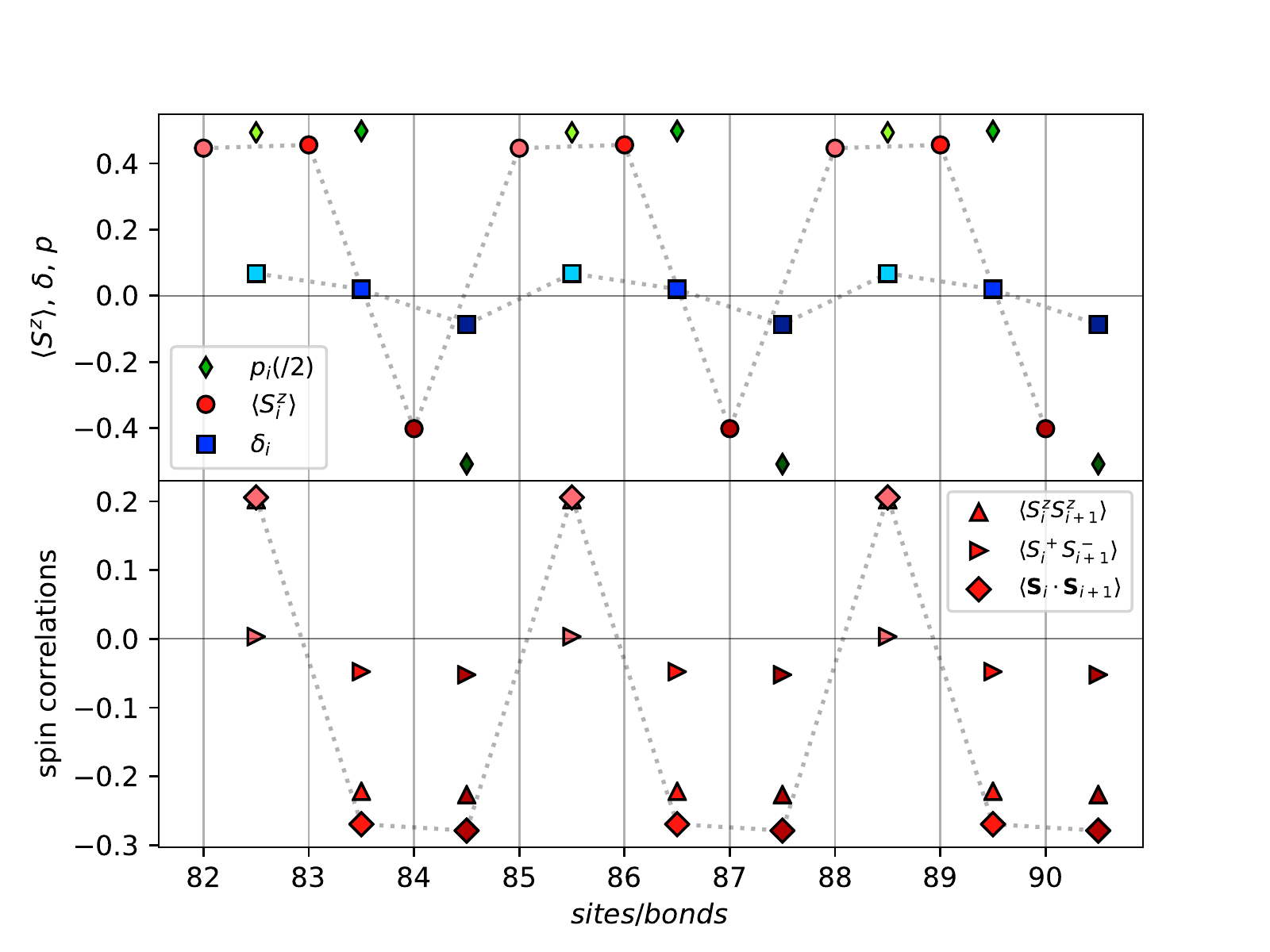}
\par\end{centering}
\caption{\label{fig: plateau profile case 4} $M=1/3$ plateau configuration
for Case 4. Markers as in Fig. 5 in the main text.}
\end{figure}
\par\end{center}

\begin{center}
\begin{figure}
\begin{centering}
\includegraphics[scale=0.55]{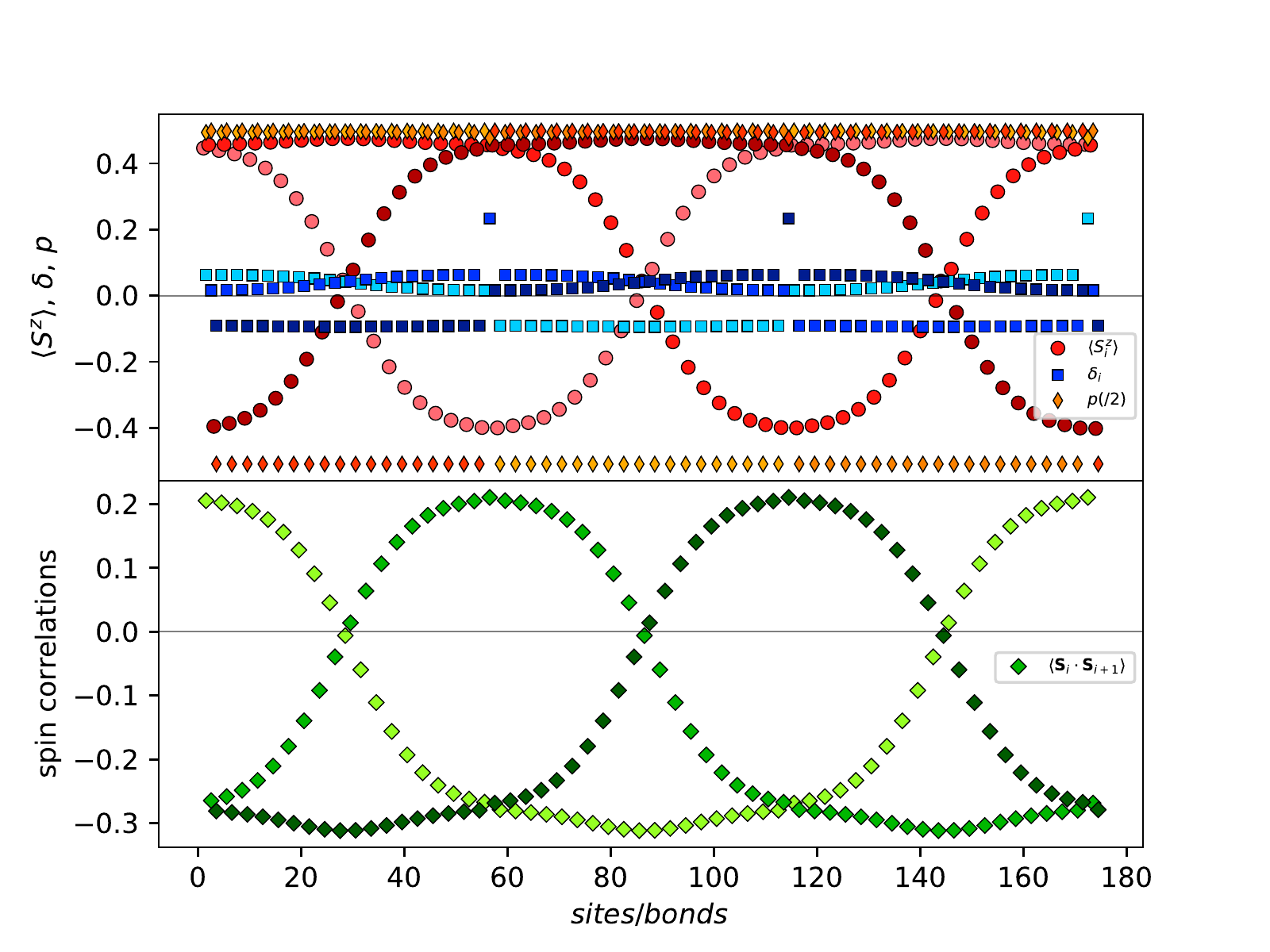}
\par\end{centering}
\caption{\label{fig: excited profile case 4} Magnetically excited state for
Case 4.}
\end{figure}
\par\end{center}